\documentclass[useAMS,usenatbib]{mn2e}
\usepackage{graphicx}
\title[Lithium and vsini in Stars with Planets]{Parent Stars of Extrasolar Planets. X. Lithium Abundances and vsini Revisited}
\author[G.\ Gonzalez et al.]{G.\ Gonzalez$^{1}$, M. K.\ Carlson$^{1}$, R. W.\ Tobin$^{2}$\\
$^{1}$Grove City College, Rockwell Hall, 100 Campus Drive, Grove City, PA 16127, USA\\
$^{2}$Department of Physics and Astronomy, Ball State University, Muncie, IN 47306 USA\\
}

\begin{document}

\date{Accepted ??. Received ??; in original form ??}

\pagerange{\pageref{firstpage}--\pageref{lastpage}} \pubyear{??}

\maketitle

\label{firstpage}

\begin{abstract}
We determine Li abundances and vsini values from new spectra of 53 stars with Doppler-detected planets not included in our previous papers in this series. We also examine two sets of stars without detected planets, which together serve as our comparison sample. Using the method of comparison of Li abundances and vsini values between two sets of stars we introduced in \citet{gg08}, we confirm that these two quantities are smaller among stars with planets compared to stars without detected planets near the solar temperature. The transition from low to high Li abundance among SWPs occurs near 5850 K, a revision of about 50 K from our previous determination. The transition from low to high vsini occurs near 6000 K, but this temperature is not as well constrained.
\end{abstract}

\section{Introduction}

In this study we revisit the observational evidence for correlations between the presence of Doppler-detected planets, Li abundance and vsini. In a previous study \citep{gg08} we confirmed the findings of some studies \citep{is04,tk05} that had indicated lower Li abundances for stars with planets (SWPs) compared to stars without detected planets over a limited range in effective temperature (T$_{\rm eff}$). We also showed that vsini and $R^{'}_{\rm HK}$ is smaller among SWPs over a similar T$_{\rm eff}$ range.

In \citet{gg08} we prepared samples of SWPs and stars without known planets with Li abundances, drawing from several diverse published studies. While these samples were heterogeneous, we corrected for small differences among the different sources of the data. The SWP and comparison samples contained of 37 and 147 stars, respectively. We introduced a new method of comparing Li abundances that takes into account differences in T$_{\rm eff}$, log g, [Fe/H] and M${\rm v}$ between stars. This approach makes it possible to search for small differences in Li abundance that are caused by processes unrelated to these four parameters.

The purpose of the present study is to test the findings of \citet{gg08} concerning the Li abundances and vsini values of SWPs. We do so with different samples of SWPs and comparison stars. In addition, our new samples are homogeneous. In Section 2 we describe our new spectroscopic observations of SWPs and our Li abudance and vsini analyses. In Section 3 we compare SWPs and stars without detected planets and discuss the results in Section 4. We present our conclusions in Section 5.

\section{Observations and analyses}

We observed a total of 53 SWPs in March and October 2008 with the McDonald Observatory 2.7-m telescope and 2dcoud\'e spectrograph. The instrument, setup and data reduction method are the same as described in our previous papers in this series (see \citet{gl00} for details). The resolving power is near 60,000, and the S/N ratio near 6700 \AA\ averages near 350 per pixel. We had not observed these SWPs in our previous observing runs. We selected them from the list of Doppler-detected dwarf SWPs with spectral types earlier than $\sim$K0 and brighter than V = 9.0 listed in the Extrasolar Planets Encyclopedia.\footnote{http://exoplanet.eu}

In addition to the new spectra we obtained in 2008, we also reanalyzed all the previous spectra of SWPs we had obtained with the McDonald 2.7-m telescope through 2002.\footnote{Unfortunately, we lost the data for 8 SWPs observed in December 1999 and described in \citet{gg01}, preventing us from reanalyzing these stars.} What's more, four SWPs from our March 2002 observing run, which is described in \citet{lg03}, are analyzed here for the first time: HD 80606, HD 149661, HD 152391 and HD 165401. Also, we analyze for the first time spectra of 24 stars without known planets and one SWP from a McDonald run in April 2002. We give additional details of the SWP and comparison star samples below.

\subsection{Measurement of equivalent widths}

We have altered our analysis procedures in the present study compared to previous ones. Previously, we had measured the equivalent widths (EWs) of the absorption lines manually; this was the most tedious step of the analysis. In the present study, we employed the program DAOSPEC, which  automates EW measurement \citep{sp08}.\footnote{We obtained DAOSPEC from Peter Stetson on November 20, 2008. This version of the program has significant enhancements over older versions.} In order to prepare each multi-order reduced spectrum for analysis with DAOSPEC, we merged its orders into a 1D spectrum and corrected for the Doppler shift. DAOSPEC identifies and measures absorption lines iteratively; it is necessary to set only a few parameters in the program.

We setup DAOSPEC to analyze each spectrum over the wavelength interval 4900 to 7950 \AA. We set the residual core flux parameter to 20\% for all the stars. We only changed two parameters for each run of DAOSPEC: the initial estimate of the FWHM of the lines and the order of the global continuum fit with a Legendre polynomial. We set the order to 1 for the warmest stars and up to 3 for the coolest metal-rich stars; higher orders were not necessary, as the input spectra had already been continuum-normalized. DAOSPEC identified between 3500 and 5000 lines in each spectrum with EWs $\ge 3$ m\AA.

DAOSPEC also requires an input linelist. Ours consists 49 Fe I, 4 Fe II, 2 C I, 1 N I, 3 O I (oxygen triplet), 2 Na I, 1 Mg I, 2 Al I, 6 Si I, 2 S I, 2 Ca I, 2 Sc II, 5 Ti I, 6 V I, 3 Cr I, 3 Co I, 6 Ni I and 1 Eu II lines. These lines are largely unblended over the range of temperatures of our target stars and are weak to moderate in strength. This linelist is very similar the one we have used in previous analyses in this series (with a few new lines added) with the following exceptions. First, a line must have a wavelength $\ge 5500$ \AA; shortward of this limit, line blending leads to inaccurate continuum placement for the cooler stars. Second, a line must not be in a region with telluric absorption lines. This is a change from our previous studies, wherein we did divide out telluric lines using a hot star spectrum. In the present study we are striving for maximum uniformity in the measurement of EWs.

DAOSPEC matches the lines in the input linelist with lines it has identified in given a spectrum. In order to pass to the next step in our analysis, the EW value of each line in our linelist determined by DAOSPEC must satisfy the following additional criteria. It must have an EW value $\ge 3$ m\AA\ and $\le 120$ m\AA, and the ''quality parameter'' value $\le 1.5$ (this parameter compares the residuals in the immediate neighborhood of a line to the overall residuals).

\subsection{Stellar atmospheric parameters}

As with our previous studies in this series, we use the program MOOG \citep{s73} with the model atmospheres of \citet{k93} to derive stellar atmospheric parameters and chemical abundances. However, for the present study we use the more recent 2002 version of MOOG.\footnote{Source code available at http://verdi.as.utexas.edu/moog.html} Given these multiple changes from our previous studies, for the sake of maximum consistency we have calculated a new set of solar-based $gf$-values using the following procedure.

First, we determined solar EWs with DAOSPEC from a spectrum of Vesta and a spectrum of the afternoon sky and averaged the results. Then, we selected 28 Fe I lines (between 5200 and 6900 \AA) with high quality EW values that appear in Table 1 of \citet{gs99}, which lists high-quality laboratory $gf$-values for Fe I lines. Next, we determined the solar abundance of Fe from each of these Fe I lines using MOOG and adjusted the microturbulence velocity parameter, $\upsilon_{\rm t}$, to minimize the dispersion; we found a best fit $\upsilon_{\rm t}$ value of 1.2 km~s$^{\rm -1}$. Then, using this value of $\upsilon_{\rm t}$, we adjusted the $gf$-values such that all the Fe I lines gave an abundance A(Fe) $= 7.47$\footnote{A(Fe) $= \log$ (N$_{\rm Fe}$/N$_{\rm H}) + 12$}; we determined the Fe II line $gf$-values using these same parameter values. Finally, we set the $gf$-values of lines from other elements so that they give the solar abundances tabulated by \citet{gs98}.

We calculated the stellar atmospheric parameters and their uncertainties using the same procedures we used in our previous papers (e.g., \citet{gv98}, \citet{gl00}). In brief, LTE is assumed and the stellar parameters are determined assuming excitation and ionisation equilibria. In particular, T$_{\rm eff}$ is determined from Fe I lines by requiring that their abundances no not display a trend with their lower excitation potentials, and surface gravity is determined by requiring that the mean Fe I and Fe II abundances be equal. The microturbulence velocity parameter is determined by requiring that the Fe I abundances do not display a trend with the reduced equivalent widths. Errors are propagated statistical uncertainties.

\subsection{Comparison stars sample}

Our comparison stars sample consists of two parts. One part is drawn from the S$^{4}$N spectroscopic survey of FGK dwarfs within 15 pc \citep{all04}, which consists of spectra obtained at two observatories: the McDonald 2.7-m telescope and the ESO 1.52-m telescope at La Silla.\footnote{We obtained the reduced S$^{4}$N survey spectra from http://hebe.as.utexas.edu/s4n/data.html} We only selected those stars observed with the McDonanld telescope; it is important to note that the spectra of \citep{all04} were obtained with the same instrument and setup that we used. In addition, stars from the S$^{4}$N survey were not included in our comparison sample if T$_{\rm eff} \le 5000$ (a few stars slightly above this cutoff were also excluded). After applying these selection criteria, we retained 36 stars from the S$^{4}$N survey.

Although the calibration of the S$^{4}$N spectra should be very close to ours, \citet{all04} did not reduce them in precisely the same way as we did ours. We compared the solar EWs from our spectra with those from S$^{4}$N and found a small difference of about 1 m\AA\ for the stronger lines (in the sense that the S$^{4}$N EWs are larger). In order to bring the two sets of EWs into agreement, we applied a simple linear transformation to the EWs from the S$^{4}$N spectra.

Four stars in our S$^{4}$N subsample are also present in our SWP sample: HD 22049, HD 69830, HD 95128 and HD 9826. These stars span nearly the entire T$_{\rm eff}$ range of our SWP sample. Their mean difference in T$_{\rm eff}$ is 11 K, and individual differences are within the estimated one-$\sigma$ uncertainties; the mean uncertainties in T$_{\rm eff}$ are also very similar. The mean difference in log g is 0.03 dex, but three stars differ between one and two-$\sigma$. The mean difference in [Fe/H] is 0.03 dex, and the individual differences are within one-$\sigma$, except for HD 22049. The differences in the Li abundances (see below) are 0.3 for HD 22049 and 0.1 dex or less for the other three stars. Therefore, we are confident that the SWP and comparison stars analysis results are on the same scales for stars warmer than about 5000 K. Following this comparison, we removed these four stars from the comparison sample.

We also analyzed for the first time spectra of 25 late-F to mid-G dwarf stars we had observed in April 2002 with the same telescope, instrument and setup we used for our SWP observations. We had selected these stars from previous spectroscopic studies of nearby stars (e.g., \citet{Cay96, egg98, Hay01}). These stars have a similar [Fe/H] distribution to our SWPs sample; they have a mean [Fe/H] value of 0.07 dex; this compares to a mean [Fe/H] of 0.12 dex for the SWPs. One of the stars from our April 2002 run, HD 170469, was discovered to host a giant planet in 2007. We added it to our SWP sample, leaving 24 comparison stars from our April 2002 run.

The S$^{4}$N subsample and our April 2002 data together form our comparison sample for the present study.\footnote{One star, HD 114710, is in common with our April 2002 sample and our S$^{4}$N subsample. Our analysis results for the two spectra were nearly identical.}. Having analyzed the old and new data in the same way, systematic differences should be insignificant.

We should note that we have not searched the comparison stars for the presence of planets. However, most of them should be on the target lists of the various planet search teams, given their brightness, spectral types and activity levels. Still, we cannot discount the possibility that one or two of our comparison stars might host an undiscovered Doppler-detectable planet.

\subsection{Lithium abundance and vsini}

We determined Li abundances and vsini for each star using spectrum synthesis with MOOG. We adopted the linelist for the Li feature near 6707 \AA\ and surrounding lines (covering about 12 \AA) from \citet{gh09}, but we left out the $^{\rm 6}$Li components, as we are not analyzing the $^{\rm 6}$Li/$^{\rm 7}$Li ratio in the present study. We adjusted the $gf$-values of some of the lines (but not Li) to produce a good match to our solar spectra. We obtained the best match for a solar Li abundance of 0.96, which is the same value obtained by \citet{gh09}.

To estimate vsini, we modeled the profiles of lines in the Li line region, including macroturbulence, $\upsilon_{\rm mac}$, instrumental broadening and limb-darkening. We adopted the same procedure as \citet{vf05} for setting the value of $\upsilon_{\rm mac}$; it depends only on T$_{\rm eff}$ according to:
\begin{eqnarray*}
\upsilon_{\rm mac} = (\upsilon_{\rm mac}(sun) + (T_{\rm eff} - 5770 K)/650 K)
\end{eqnarray*}

We calibrated this equation with our solar spectra by setting vsini equal to 1.6 km~s$^{\rm -1}$; from this, we determined $\upsilon_{\rm mac}(sun) = 4.1$ km~s$^{\rm -1}$. We estimate the uncertainty of $\upsilon_{\rm mac}$ calculated from this equation to be $\pm$ 0.1 km~s$^{\rm -1}$. 

We determined vsini for each star by comparing the observed spectrum to synthetic spectra over the region 6703 to 6706 \AA. Specifically, after setting the values of the instrumental broadening and $\upsilon_{\rm mac}$ and choosing initial values of vsini, Fe abundance and continuum level, we compared the synthetic and observed spectra in the wavelength regions 6703.4 to 6703.8 \AA\ and 6704.9 to 6705.3 \AA. We adjusted (manually) vsini, Fe abundance and continuum level until the differences between the observed and synthetic spectra in these regions were comparable to the unmodeled variations in the surrounding spectrum. The typical uncertainty of our vsini estimates is about $\pm$ 0.5 km~s$^{\rm -1}$; uncertainty estimates are based on the uncertainty of $\upsilon_{\rm mac}$ quoted above and on the size of the residuals of the fit compared to the observed spectrum. We show sample syntheses of the 6700 \AA\ spectral region in Figure 1.

We estimate the uncertainty of the Li abundance due to noise and unmodelled lines in a typical observed spectrum to be about $\pm$ 0.05 dex (see Figure 1b). The total uncertainty of the Li abundance is based on the quadrature sum of this estimate and the uncertainty of Li due to the uncertainty of T$_{\rm eff}$. The results of our spectroscopic analyses of the 53 new SWPs are listed in Table 1. The new results for the previously analyzed SWPs are listed in Table 2, and the results for the comparison stars are in Table 3. We will present our abundance results for elements other than Li and Fe in a separate paper.

\begin{figure}
  \includegraphics[width=3.5in]{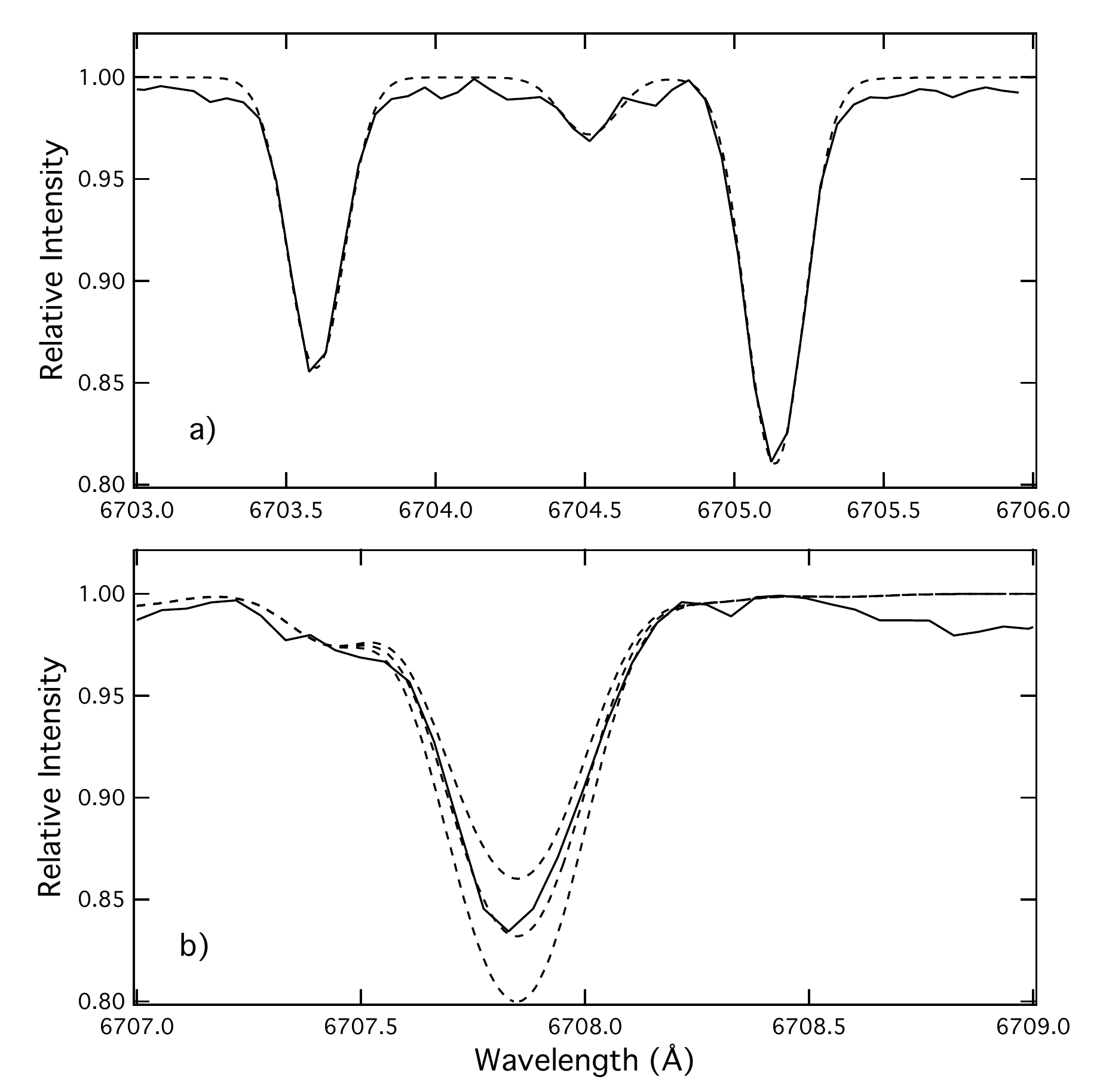}
 \caption{The observed spectrum of HD 74156 is shown as a solid curve in each panel. Best-fit synthetic spectrum of the region containing the two Fe I lines used to determine vsini is shown as a dashed curve (panel a). Best-fit and $\pm$ 0.1 dex Li abundance syntheses are shown as dashed curves (panel b).}
\end{figure}

\begin{figure}
  \includegraphics[width=3.5in]{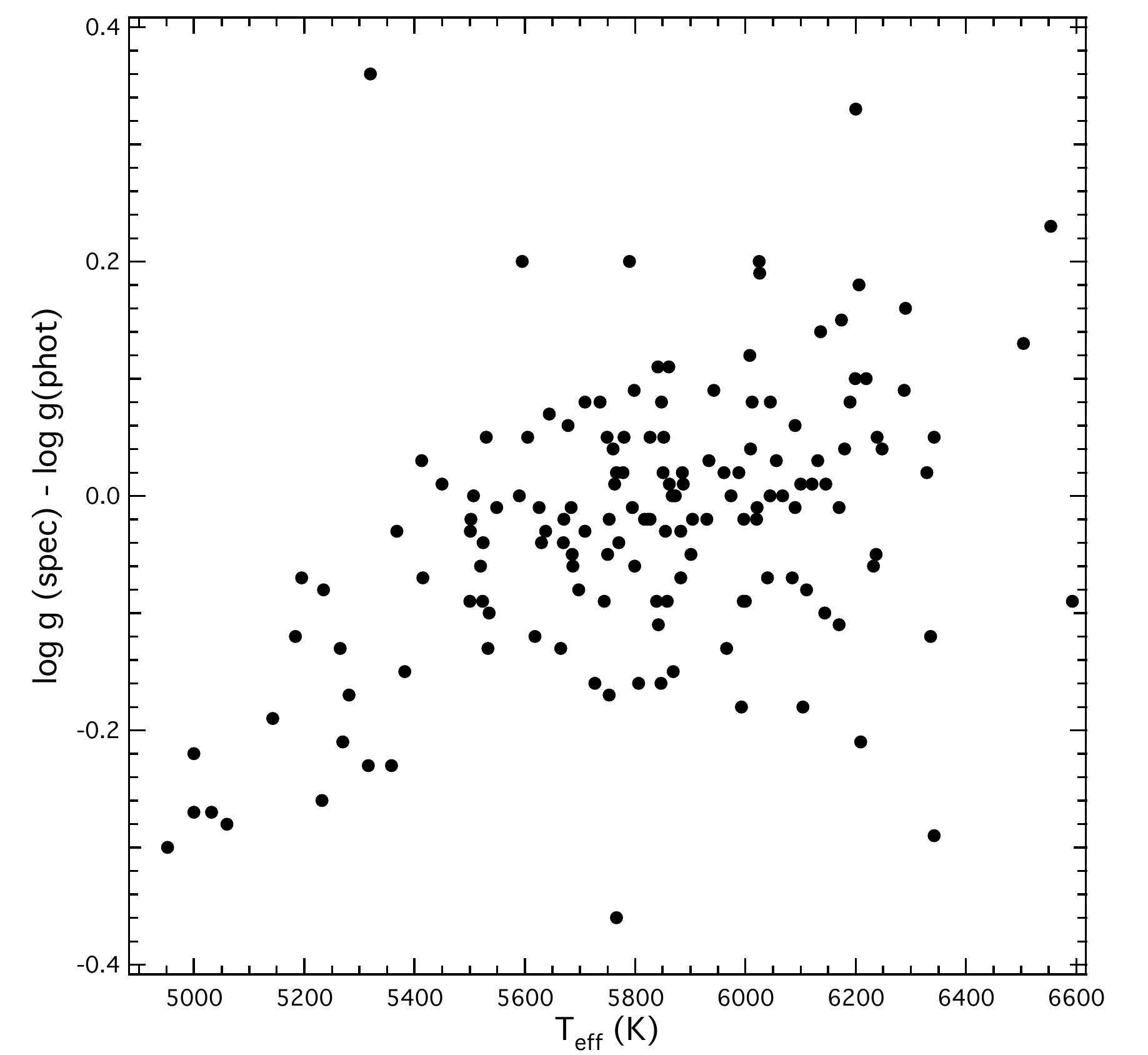}
 \caption{Differences between the spectroscopic and photometric log g values for all the stars analyzed in the present work.}
\end{figure}

\begin{table*}
\centering
\begin{minipage}{160mm}
\caption{Parameters determined from our spectroscopic analyses of 53 SWPs in columns 3 to 8. Derived parameters based on stellar isochrones are given in columns 9 to 11.}
\label{xmm}
\begin{tabular}{llrcccccccc}
\hline
Star & & T$_{\rm eff}$ & log g & $\zeta_{\rm t}$ (km~s$^{\rm -1}$) & [Fe/H] & log $\epsilon$(Li) & vsini (km~s$^{\rm -1}$) & mass (M$_{\odot}$) & log g & age (Gyr)\\
HD & HIP & $\pm$ & $\pm$ & $\pm$ & $\pm$ & $\pm$ & $\pm$ & $\pm$ & $\pm$ & $\pm$\\
\hline
3651 & 3093 & 5143 & 4.30 & 0.8 & ~0.184 & ~~-0.30 & ~~1.6 & 0.89 & 4.49 & ~5.3\\
 & & 38 & 0.07 & ~0.09 & ~0.024 & ~~~0.05 & ~~0.5 & 0.03 & 0.02 & ~4.5\\
4113 & 3391 & 5626 & 4.30 & 0.9 & ~0.240 & ~~~0.53 & ~~3.5 & 1.00 & 4.31 & ~6.6\\
 & & 31 & 0.06 & ~0.08 & ~0.022 & ~~~0.06 & ~~0.4 & 0.03 & 0.05 & ~3.1\\
11506 & 8770 & 6136 & 4.42 & 1.5 & ~0.323 & ~~~2.71 & ~~6.0 & 1.21 & 4.28 & ~1.6\\
 & & 50 & 0.06 & ~0.09 & ~0.036 & ~~~0.06 & ~~0.5 & 0.02 & 0.03 & ~1.0\\
 11964 & 9094 & 5265 & 3.74 & 1.0 & ~0.121 & ~~~1.19 & ~~3.3 & 1.10 & 3.87 & ~7.6\\
 & & 30 & 0.05 & ~0.07 & ~0.020 & ~~~0.06 & ~~0.5 & 0.03 & 0.03 & ~0.6\\
   16175 & 12191 & 5930 & 4.02 & 1.4 & ~0.310 & ~~~2.45 & ~~5.5 & 1.25 & 4.04 & ~3.8\\
 & & 38 & 0.05 & ~0.05 & ~0.029 & ~~~0.06 & ~~0.5 & 0.05 & 0.04 & ~0.7\\
  17156 & 13192 & 5993 & 4.04 & 1.4 & ~0.181 & ~~~2.40 & ~~5.0 & 1.16 & 4.22 & ~3.5\\
 & & 47 & 0.03 & ~0.07 & ~0.035 & ~~~0.06 & ~~0.8 & 0.03 & 0.05 & ~1.2\\
    20367 & 15323 & 6090 & 4.34 & 1.4 & ~0.105 & ~~~2.67 & $<$3.3 & 1.14 & 4.35 & ~1.3\\
 & & 40 & 0.04 & ~0.10 & ~0.030 & ~~~0.06 & ~~0.2 & 0.03 & 0.03 & ~1.1\\
   20782 & 15527 & 5760 & 4.36 & 1.2 & -0.048 & ~~~0.64 & ~~3.0 & 0.96 & 4.32 & ~7.9\\
 & & 26 & 0.04 & ~0.09 & ~0.019 & ~~~0.07 & ~~0.6 & 0.03 & 0.04 & ~3.1\\
   23596 & 17747 & 6045 & 4.18 & 1.2 & ~0.286 & ~~~2.77 & ~~4.2 & 1.22 & 4.18 & ~3.1\\
 & & 56 & 0.14 & ~0.07 & ~0.042 & ~~~0.08 & ~~0.6 & 0.02 & 0.04 & ~0.5\\
   27063 & 19925 & 5750 & 4.39 & 1.2 & ~0.065 & ~~~1.62 & $<$2.0 & 1.00 & 4.44 & ~2.2\\
 & & 20 & 0.02 & ~0.06 & ~0.020 & ~~~0.05 & ~~0.7 & 0.03 & 0.03 & ~2.1\\
   33283 & 23889 & 5988 & 4.00 & 1.5 & ~0.309 & ~~~2.16 & ~~5.0 & 1.34 & 3.98 & ~3.2\\
 & & 21 & 0.04 & ~0.03 & ~0.016 & ~~~0.05 & ~~0.7 & 0.06 & 0.05 & ~0.5\\
  33564 & 25110 & 6554 & 4.49 & 2.0 & ~0.211 & ~~~2.17 & ~14.5 & 1.34 & 4.26 & ~0.5\\
 & & 93 & 0.05 & ~0.2 & ~0.063 & ~~~0.09 & ~~0.5 & 0.02 & 0.02 & ~0.3\\
  37605 & 26664 & 5270 & 4.26 & 1.0 & ~0.265 & $<$-0.12 & ~~3.5 & 0.91 & 4.47 & ~5.1\\
 & & 47 & 0.07 & ~0.07 & ~0.033 & ~~~0.05 & ~~0.7 & 0.03 & 0.03 & ~4.2\\
  43691 & 30057 & 6144 & 4.05 & 1.5 & ~0.268 & ~~~1.98 & ~~6.0 & 1.27 & 4.15 & ~2.5\\
 & & 34 & 0.05 & ~0.07 & ~0.026 & ~~~0.06 & ~~1.0 & 0.04 & 0.06 & ~0.5\\
  45350 & 30860 & 5523 & 4.10 & 1.0 & ~0.292 & ~$<$0.01 & ~~1.6 & 0.98 & 4.19 & ~10.1\\
 & & 30 & 0.04 & ~0.05 & ~0.023 & ~~~0.05 & ~~0.6 & 0.02 & 0.04 & ~1.2\\
  45652 & 30905 & 5232 & 4.20 & 0.9 & ~0.286 & $<$-0.64 & ~~3.5 & 0.91 & 4.46 & ~6.2\\
 & & 48 & 0.07 & ~0.06 & ~0.032 & ~~~0.08 & ~~0.5 & 0.03 & 0.03 & ~4.4\\
  47186 & 31540 & 5630 & 4.27 & 1.1 & ~0.252 & $<$-0.40 & ~~4.0 & 1.00 & 4.31 & ~6.9\\
 & & 38 & 0.06 & ~0.06 & ~0.030 & ~~~0.06 & ~~0.5 & 0.03 & 0.04 & ~2.8\\
  49674 & 32916 & 5590 & 4.38 & 1.0 & ~0.325 & ~~~0.39 & ~~0.9 & 0.99 & 4.38 & ~4.9\\
 & & 31 & 0.03 & ~0.08 & ~0.021 & ~~~0.05 & ~~0.5 & 0.03 & 0.05 & ~3.7\\
  50499 & 32970 & 6045 & 4.31 & 1.5 & ~0.312 & ~~~2.61 & ~~4.2 & 1.19 & 4.23 & ~2.9\\
 & & 27 & 0.03 & ~0.05 & ~0.020 & ~~~0.06 & ~~0.8 & 0.02 & 0.03 & ~0.9\\
  60532 & 36795 & 6121 & 3.73 & 2.2 & -0.250 & ~~~1.63 & ~~8.0 & 1.37 & 3.72 & ~2.8\\
 & & 85 & 0.05 & ~0.25 & ~0.060 & ~~~0.08 & ~~0.6 & 0.04 & 0.03 & ~0.2\\
  66428 & 39417 & 5686 & 4.26 & 1.0 & ~0.324 & ~$<$0.58 & $<$0.5 & 1.02 & 4.31 & ~5.5\\
 & & 35 & 0.05 & ~0.06 & ~0.026 & ~~~0.09 & ~~0.4 & 0.03 & 0.06 & ~3.1\\
  69830 & 40693 & 5413 & 4.52 & 1.0 & -0.006 & ~~~0.65 & $<$0.5 & 0.90 & 4.49 & ~4.4\\
 & & 25 & 0.07 & ~0.11 & ~0.019 & ~~~0.05 & ~~0.5 & 0.03 & 0.02 & ~3.9\\
  70573 & & 5807 & 4.35 & 1.8 & -0.049 & ~~~2.75 & 13.5 & & & \\
 & & 85 & 0.08 & ~0.16 & ~0.064 & ~~~0.09 & ~~0.5 & & & \\
  70642 & 40952 & 5671 & 4.40 & 1.0 & ~0.220 & ~~~0.75 & ~~0.5 & 1.01 & 4.42 & ~2.4\\
 & & 31 & 0.04 & ~0.09 & ~0.021 & ~~~0.09 & ~~0.5 & 0.03 & 0.03 & ~2.3\\
  72659 & 42030 & 5883 & 4.08 & 1.3 & -0.022 & ~~~2.19 & $<$2.2 & 1.05 & 4.15 & ~7.0\\
 & & 37 & 0.08 & ~0.07 & ~0.029 & ~~~0.06 & ~~0.5 & 0.03 & 0.04 & ~1.2\\
  74156 & 42723 & 6009 & 4.12 & 1.5 & ~0.064 & ~~~2.41 & ~~4.2 & 1.19 & 4.08 & ~4.3\\
 & & 40 & 0.04 & ~0.08 & ~0.030 & ~~~0.06 & ~~0.5 & 0.04 & 0.04 & ~0.8\\
  75898 & 43674 & 5966 & 3.98 & 1.3 & ~0.249 & ~~~2.51 & ~~5.0 & 1.22 & 4.11 & ~3.8\\
 & & 55 & 0.07 & ~0.08 & ~0.042 & ~~~0.07 & ~~0.5 & 0.05 & 0.06 & ~0.7\\
  80606 & 45982 & 5507 & 4.33 & 1.0 & ~0.339 & ~$<$0.17 & ~~2.0 & 0.98 & 4.33 & ~5.5\\
 & & 32 & 0.07 & ~0.06 & ~0.022 & ~~~0.07 & ~~0.4 & 0.06 & 0.15 & ~4.6\\
  81040 & 46076 & 5798 & 4.55 & 1.2 & -0.033 & ~~~1.92 & ~~3.0 & 0.97 & 4.46 & ~2.1\\
 & & 26 & 0.03 & ~0.07 & ~0.020 & ~~~0.06 & ~~0.7 & 0.03 & 0.02 & ~2.0\\
  86081 & 48711 & 6000 & 4.16 & 1.6 & ~0.183 & ~~~1.85 & ~~5.0 & 1.16 & 4.25 & ~2.9\\
 & & 27 & 0.04 & ~0.04 & ~0.020 & ~~~0.05 & ~~0.6 & 0.04 & 0.07 & ~1.6\\
\hline
\end{tabular}
\end{minipage}
\end{table*}

\begin{table*}
\centering
\contcaption{}
\begin{tabular}{llrcccccccc}
\hline
Star & & T$_{\rm eff}$ & log g & $\zeta_{\rm t}$ (km~s$^{\rm -1}$) & [Fe/H] & log $\epsilon$(Li) & vsini (km~s$^{\rm -1}$) & mass (M$_{\odot}$) & log g & age (Gyr)\\
HD & HIP & $\pm$ & $\pm$ & $\pm$ & $\pm$ & $\pm$ & $\pm$ & $\pm$ & $\pm$ & $\pm$\\
\hline
  88133 & 49813 & 5320 & 3.69 & 1.1 & ~0.325 & ~~~1.56 & ~~2.2 & 1.97 & 3.33 & ~1.2\\
 & & 36 & 0.05 & ~0.06 & ~0.024 & ~~~0.07 & ~~0.8 & 0.10 & 0.06 & ~0.2\\
 89307 & 50473 & 5961 & 4.37 & 1.5 & -0.120 & ~~~2.10 & ~~2.9 & 1.01 & 4.35 & ~4.9\\
 & & 34 & 0.03 & ~0.07 & ~0.024 & ~~~0.06 & ~~0.5 & 0.04 & 0.04 & ~2.9\\
  107148 & 60081 & 5763 & 4.32 & 1.1 & ~0.316 & ~~~1.06 & ~~0.7 & 1.05 & 4.31 & 4.6\\
 & & 13 & 0.03 & ~0.04 & ~0.010 & ~~~0.05 & ~~0.5 & 0.03 & 0.05 & ~2.5\\
  114729 & 64459 & 5778 & 4.04 & 1.6 & -0.320 & ~~~1.73 & ~~2.3 & 0.94 & 4.02 & 10.9\\
 & & 31 & 0.06 & ~0.10 & ~0.010 & ~~~0.06 & ~~0.6 & 0.01 & 0.03 & ~0.5\\
  114762 & 64426 & 5766 & 3.87 & 1.6 & -0.779 & ~~~1.85 & $<$1.8 & 0.82 & 4.23 & 12.6\\
 & & 51 & 0.04 & ~0.22 & ~0.040 & ~~~0.07 & ~~0.4 & 0.02 & 0.15 & ~0.1\\
   117207 & 65808 & 5618 & 4.19 & 1.0 & ~0.259 & ~$<$0.03 & $<$1.1 & 0.99 & 4.31 & 7.1\\
 & & 30 & 0.04 & ~0.07 & ~0.023 & ~~~0.05 & ~~0.5 & 0.03 & 0.04 & ~2.8\\
  118203 & 66192 & 5827 & 3.95 & 1.4 & ~0.209 & ~~~2.47 & ~~5.0 & 1.26 & 3.97 & 4.0\\
 & & 25 & 0.03 & ~0.05 & ~0.020 & ~~~0.05 & ~~0.6 & 0.06 & 0.05 & ~0.7\\
  132406 & 73146 & 5727 & 4.10 & 1.3 & ~0.111 & ~~~1.09 & ~~4.0 & 1.01 & 4.26 & 7.4\\
 & & 32 & 0.03 & ~0.08 & ~0.024 & ~~~0.06 & ~~0.7 & 0.03 & 0.05 & ~2.2\\
  136118 & 74948 & 6248 & 4.21 & 1.8 & -0.016 & ~~~2.18 & ~~7.5 & 1.21 & 4.17 & ~3.1\\
 & & 45 & 0.03 & ~0.09 & ~0.031 & ~~~0.06 & ~~0.5 & 0.02 & 0.03 & ~0.5\\
  141937 & 77740 & 5855 & 4.37 & 1.2 & ~0.119 & ~~~2.21 & $<$1.9 & 1.05 & 4.40 & ~1.8\\
 & & 20 & 0.03 & ~0.06 & ~0.016 & ~~~0.05 & ~~0.5 & 0.03 & 0.03 & ~1.7\\
   147513 & 80337 & 5886 & 4.46 & 1.2 & ~0.073 & ~~~1.89 & $<$1.7 & 1.03 & 4.44 & ~1.4\\
 & & 22 & 0.03 & ~0.06 & ~0.017 & ~~~0.05 & ~~0.7 & 0.02 & 0.02 & ~1.2\\
  149026 & 8038 & 6131 & 4.22 & 1.5 & ~0.307 & ~~~2.31 & ~~6.0 & 1.24 & 4.19 & ~2.4\\
 & & 35 & 0.05 & ~0.06 & ~0.026 & ~~~0.06 & ~~0.5 & 0.03 & 0.05 & ~0.7\\
  149661 & 81300 & 5184 & 4.41 & 0.8 & ~0.051 & ~~-0.20 & $<$2.2 & 0.86 & 4.53 & ~4.5\\
 & & 39 & 0.06 & ~0.08 & ~0.024 & ~~~0.05 & ~~0.3 & 0.02 & 0.02 & ~4.1\\
  150706 & 80902 & 5883 & 4.39 & 1.2 & -0.041 & ~~~2.46 & ~~4.5 & 1.01 & 4.42 & ~2.3\\
 & & 33 & 0.05 & ~0.07 & ~0.025 & ~~~0.06 & ~~0.6 & 0.04 & 0.03 & ~2.1\\
  152391 & 82588 & 5450 & 4.51 & 1.2 & ~0.003 & ~~~1.11 & ~~3.8 & 0.90 & 4.50 & ~3.9\\
 & & 31 & 0.07 & ~0.07 & ~0.022 & ~~~0.06 & ~~0.5 & 0.03 & 0.02 & ~3.5\\
   154345 & 83389 & 5358 & 4.26 & 0.8 & -0.105 & $<-0.29$ & $<$1.5 & 0.88 & 4.49 & ~6.1\\
 & & 51 & 0.08 & ~0.10 & ~0.038 & ~~~0.05 & ~~0.3 & 0.03 & 0.02 & ~4.7\\
  156846 & 84856 & 6068 & 3.98 & 1.5 & ~0.196 & ~~~1.07 & ~~5.0 & 1.37 & 3.98 & ~2.8\\
 & & 33 & 0.07 & ~0.06 & ~0.024 & ~~~0.05 & ~~0.4 & 0.04 & 0.03 & ~0.4\\
  164922 & 88348 & 5281 & 4.25 & 0.8 & ~0.205 & ~$<$0.23 & $<$2.0 & 0.91 & 4.42 & ~7.6\\
 & & 36 & 0.07 & ~0.09 & ~0.026 & ~~~0.08 & ~~0.5 & 0.03 & 0.03 & ~4.0\\
  165401 & 88622 & 5790 & 4.62 & 1.5 & -0.416 & ~~~0.75 & ~~4.5 & 0.86 & 4.42 & ~8.6\\
 & & 31 & 0.05 & ~0.12 & ~0.023 & ~~~0.07 & ~~0.5 & 0.02 & 0.03 & ~3.0\\
  168746 & 90004 & 5549 & 4.35 & 1.1 & -0.093 & ~~~0.72 & $<$0.5 & 0.91 & 4.36 & 10.0\\
 & & 26 & 0.03 & ~0.06 & ~0.022 & ~~~0.05 & ~~0.5 & 0.02 & 0.03 & ~1.9\\
  183263 & 95740 & 5943 & 4.37 & 1.4 & ~0.300 & ~~~2.29 & ~~1.8 & 1.13 & 4.28 & ~3.0\\
 & & 35 & 0.04 & ~0.06 & ~0.026 & ~~~0.06 & ~~0.4 & 0.03 & 0.05 & ~1.7\\
  185269 & 96507 & 5997 & 3.94 & 1.7 & ~0.101 & ~~~2.21 & ~~6.0 & 1.32 & 3.96 & ~3.4\\
 & & 47 & 0.03 & ~0.07 & ~0.036 & ~~~0.06 & ~~0.5 & 0.05 & 0.03 & ~0.5\\
  188015 & 97769 & 5684 & 4.26 & 1.0 & ~0.275 & $<$-0.10 & $<$0.5 & 1.02 & 4.27 & ~6.6\\
 & & 31 & 0.08 & ~0.07 & ~0.025 & ~~~0.05 & ~~0.5 & 0.03 & 0.05 & ~2.4\\
  189733 & 98505 & 4952 & 4.26 & 0.9 & ~0.010 & $<$-0.30 & ~~4.5 & 0.81 & 4.56 & ~4.8\\
 & & 64 & 0.12 & ~0.08 & ~0.036 & ~~~0.14 & ~~0.5 & 0.02 & 0.02 & ~4.0\\
  190360 & 98767 & 5500 & 4.19 & 1.0 & ~0.227 & ~~~0.41 & ~~2.5 & 0.96 & 4.28 & ~9.6\\
 & & 36 & 0.06 & ~0.10 & ~0.028 & ~~~0.06 & ~~0.5 & 0.02 & 0.03 & ~1.7\\
  196885 & 101966 & 6288 & 4.36 & 1.8 & ~0.210 & ~~~2.58 & ~~8.0 & 1.26 & 4.27 & ~1.0\\
 & & 61 & 0.04 & ~0.10 & ~0.043 & ~~~0.06 & ~~0.5 & 0.02 & 0.03 & ~0.7\\
  208487 & 108375 & 6200 & 4.67 & 1.8 & ~0.085 & ~~~2.51 & ~~4.5 & 1.16 & 4.34 & ~1.0\\
 & & 65 & 0.06 & ~0.14 & ~0.047 & ~~~0.07 & ~~0.5 & 0.03 & 0.03 & ~0.9\\
  216770 & 113238 & 5316 & 4.22 & 0.95 & ~0.293 & ~~~0.62 & ~~1.2 & 0.92 & 4.45 & ~5.6\\
 & & 47 & 0.09 & ~0.09 & ~0.033 & ~~~0.10 & ~~0.4 & 0.03 & 0.03 & ~4.3\\
  219828 & 115100 & 5861 & 4.21 & 1.5 & ~0.175 & ~~~2.17 & ~~4.5 & 1.16 & 4.1 & ~4.9\\
 & & 38 & 0.03 & ~0.08 & ~0.029 & ~~~0.06 & ~~0.5 & 0.05 & 0.05 & ~0.9\\
  224693 & 118319 & 6024 & 4.23 & 1.6 & ~0.270 & ~~~1.89 & ~~5.5 & 1.31 & 4.03 & ~3.1\\
 & & 41 & 0.04 & ~0.07 & ~0.031 & ~~~0.06 & ~~0.5 & 0.06 & 0.07 & ~0.5\\
   & 14810 & 5501 & 4.32 & 1.0 & ~0.274 & ~0.92 & ~$<$1.5 & 0.95 & 4.35 & ~7.2\\
 & & 38 & 0.05 & ~0.10 & ~0.027 & ~~~0.06 & ~~0.5 & 0.03 & 0.06 & ~4.6\\
\hline
\end{tabular}
\end{table*}

\begin{table*}
\centering
\begin{minipage}{160mm}
\caption{Parameters determined from our spectroscopic reanalyses of the SWPs included in our previous papers. Note: the two spectra of HD 8574 were obtained on different runs. HD 170469 is from the April 2002 run. Columns are the same as in Table 1.}
\label{xmm}
\begin{tabular}{llccccccccc}
\hline
Star & & T$_{\rm eff}$ & log g & $\zeta_{\rm t}$ (km~s$^{\rm -1}$) & [Fe/H] & log $\epsilon$(Li) & vsini (km~s$^{\rm -1}$) & mass (M$_{\odot}$) & log g & age (Gyr)\\
HD & HIP & $\pm$ & $\pm$ & $\pm$ & $\pm$ & $\pm$ & $\pm$ & $\pm$ & $\pm$ & $\pm$\\
\hline
4203 & 3502 & 5535 & 4.04 & ~~1.1 & ~0.371 & ~$<$0.50 & ~~1.7 & 1.00 & 4.14 & ~9.8\\
 & & ~~~30 & 0.04 & 0.04 & ~0.023 & ~~~0.06 & ~~0.4 & 0.03 & 0.07 & ~1.2\\
4208 & 3479 & 5638 & 4.46 & ~~1.1 & -0.232 & ~$<$0.44 & $<$0.5 & 0.89 & 4.49 & ~4.7\\
 & & ~~~25 & 0.04 & 0.14 & ~0.020 & ~~~0.07 & ~~0.5 & 0.03 & 0.03 & ~4.0\\
6434 & 5054 & 5744 & 4.29 & ~~1.4 & -0.588 & ~~~0.70 & $<$1.0 & 0.84 & 4.38 & 11.3\\
 & & ~~~45 & 0.03 & 0.24 & ~0.034 & ~~~0.07 & ~~0.5 & 0.01 & 0.01 & ~0.3\\
8574(a) & 6643 & 6040 & 4.10 & ~~1.5 & -0.040 & ~~~2.48 & ~~4.5 & 1.10 & 4.17 & ~5.3\\
 & & ~~~33 & 0.02 & 0.07 & ~0.026 & ~~~0.06 & ~~0.5 & 0.03 & 0.04 & ~1.0\\
8574(b) & 6643 & 6090 & 4.26 & ~~1.7 & -0.017 & ~~~2.43 & ~~4.5 & 1.13 & 4.20 & ~4.4\\
 & & ~~~35 & 0.04 & 0.10 & ~0.025 & ~~~0.06 & ~~0.5 & 0.03 & 0.04 & ~0.9\\
9826 & 7513 & 6170 & 4.00 & ~~1.8 & ~0.084 & ~~~2.16 & ~~9.5 & 1.26 & 4.11 & ~3.1\\
 & & ~~~48 & 0.08 & 0.10 & ~0.037 & ~~~0.06 & ~~0.4 & 0.03 & 0.03 & ~0.5\\
12661 & 9683 & 5670 & 4.31 & ~~1.0 & ~0.376 & ~~~1.92 & ~~1.2 & 1.01 & 4.35 & ~4.6\\
 & & ~~~27 & 0.03 & 0.06 & ~0.021 & ~~~0.06 & ~~0.4 & 0.03 & 0.04 & ~3.1\\
16141 & 12048 & 5770 & 4.06 & ~~1.2 & ~0.163 & ~~~1.34 & ~~1.9 & 1.10 & 4.10 & ~6.2\\
 & & ~~~25 & 0.04 & 0.05 & ~0.019 & ~~~0.06 & ~~0.8 & 0.03 & 0.04 & ~1.1\\
19994 & 14954 & 6190 & 4.16 & ~~1.8 & ~0.207 & ~~~1.83 & ~~8.0 & 1.33 & 4.08 & ~2.5\\
 & & ~~~45 & 0.04 & 0.08 & ~0.032 & ~~~0.07 & ~~0.3 & 0.01 & 0.03 & ~0.3\\
22049 & 16537 & 5000 & 4.30 & ~~0.9 & -0.088 & ~~~0.36 & ~~3.2 & 0.80 & 4.57 & ~4.8\\
 & & ~~~41 & 0.09 & 0.07 & ~0.024 & ~~~0.07 & ~~0.4 & 0.02 & 0.02 & ~3.9\\
28185 & 20723 & 5605 & 4.36 & ~~1.0 & ~0.233 & ~~~0.63 & ~~2.0 & 0.99 & 4.31 & ~7.3\\
 & & ~~~32 & 0.05 & 0.06 & ~0.022 & ~~~0.07 & ~~0.3 & 0.03 & 0.05 & ~3.0\\
37124 & 26381 & 5520 & 4.30 & ~~1.0 & ~0.228 & ~~~0.35 & $<$2.0 & 0.92 & 4.36 & 10.2\\
 & & ~~~40 & 0.09 & 0.14 & ~0.023 & ~~~0.06 & ~~0.4 & 0.01 & 0.02 & ~1.3\\
46375 & 31246 & 5195 & 4.34 & ~~0.8 & ~0.297 & ~~~0.02 & $<$1.5 & 0.90 & 4.41 & ~9.8\\
 & & ~~~36 & 0.05 & 0.09 & ~0.021 & ~~~0.06 & ~~0.6 & 0.01 & 0.02 & ~2.1\\
50554 & 33212 & 6056 & 4.42 & ~~1.4 & ~0.006 & ~~~2.44 & ~~3.5 & 1.09 & 4.39 & ~1.4\\
 & & ~~~28 & 0.02 & 0.07 & ~0.006 & ~~~0.02 & ~~0.5 & 0.03 & 0.03 & ~1.2\\
68988 & 40687 & 5901 & 4.32 & ~~1.2 & ~0.334 & ~~~2.02 & ~~3.0 & 1.10 & 4.37 & ~4.5\\
 & & ~~~34 & 0.03 & 0.06 & ~0.025 & ~~~0.06 & ~~0.4 & 0.03 & 0.03 & ~1.5\\
75289 & 43177 & 6174 & 4.46 & ~~1.5 & ~0.290 & ~~~2.66 & ~~4.2 & 1.20 & 4.31 & ~1.0\\
 & & ~~~30 & 0.08 & 0.05 & ~0.022 & ~~~0.06 & ~~0.4 & 0.02 & 0.02 & ~0.7\\
80606 & 45982 & 5507 & 4.33 & ~~1.0 & ~0.339 & ~$<$0.17 & ~~2.0 & 0.98 & 4.33 & ~5.5\\
 & & ~~~32 & 0.07 & 0.06 & ~0.022 & ~~~0.07 & ~~0.4 & 0.06 & 0.15 & ~4.6\\
82943 & 47007 & 6012 & 4.43 & ~~1.4 & ~0.295 & ~~~2.39 & $<$1.5 & 1.14 & 4.35 & ~1.2\\
 & & ~~~35 & 0.04 & 0.10 & ~0.025 & ~~~0.06 & ~~0.4 & 0.02 & 0.02 & ~1.0\\
89744 & 50786 & 6237 & 3.88 & ~~2.0 & ~0.170 & ~~~1.88 & ~~9.0 & 1.48 & 3.93 & ~2.2\\
 & & ~~~50 & 0.04 & 0.10 & ~0.035 & ~~~0.07 & ~~0.5 & 0.02 & 0.03 & ~0.2\\
95128 & 53721 & 5867 & 4.26 & ~~1.3 & ~0.025 & ~~~1.66 & ~~2.9 & 1.03 & 4.26 & ~6.5\\
 & & ~~~21 & 0.02 & 0.06 & ~0.016 & ~~~0.05 & ~~0.5 & 0.03 & 0.02 & ~1.6\\
106252 & 59610 & 5839 & 4.23 & ~~1.2 & -0.083 & ~~~1.53 & $<$1.5 & 0.97 & 4.32 & ~7.2\\
 & & ~~~31 & 0.05 & 0.06 & ~0.023 & ~~~0.06 & ~~0.6 & 0.03 & 0.04 & ~3.0\\
114762 & 64426 & 5766 & 3.87 & ~~1.6 & -0.779 & ~~~1.85 & $<$1.8 & 0.82 & 4.23 & 12.6\\
 & & ~~~51 & 0.04 & 0.22 & ~0.040 & ~~~0.07 & ~~0.4 & 0.02 & 0.15 & ~0.1\\
114783 & 64457 & 5000 & 4.32 & ~~0.8 & ~0.090 & ~$<$0.01 & $<$1.0 & 0.84 & 4.54 & ~5.2\\
 & & ~~~57 & 0.22 & 0.10 & ~0.021 & ~~~0.05 & ~~0.4 & 0.02 & 0.02 & ~4.3\\
117176 & 65721 & 5502 & 3.88 & ~~1.2 & -0.070 & ~~~1.78 & ~~2.7 & 1.07 & 3.90 & ~8.1\\
 & & ~~~17 & 0.02 & 0.04 & ~0.013 & ~~~0.05 & ~~0.4 & 0.01 & 0.03 & ~0.3\\
120136 & 67275 & 6342 & 3.92 & ~~1.8 & ~0.195 & ~~~2.14 & ~13.5 & 1.29 & 4.21 & ~1.6\\
 & & ~~100 & 0.10 & 0.2 & ~0.070 & ~~~0.10 & ~~0.5 & 0.02 & 0.03 & ~0.6\\
130322 & 72339 & 5368 & 4.46 & ~~0.9 & ~0.080 & ~~~0.61 & $<$1.7 & 0.90 & 4.49 & ~4.2\\
 & & ~~31 & 0.05 & 0.08 & ~0.020 & ~~~0.06 & ~~0.3 & 0.03 & 0.03 & ~3.8\\
134987 & 74500 & 5687 & 4.17 & ~~1.1 & ~0.299 & ~~~0.94 & $<$2.0 & 1.03 & 4.23 & ~7.4\\
 & & ~~35 & 0.05 & 0.05 & ~0.026 & ~~~0.06 & ~~0.2 & 0.02 & 0.04 & ~1.5\\
136118 & 74948 & 6248 & 4.21 & ~~1.8 & -0.016 & ~~~2.18 & ~~7.5 & 1.21 & 4.17 & ~3.1\\
 & & ~~~45 & 0.03 & 0.09 & ~0.031 & ~~~0.06 & ~~0.5 & 0.02 & 0.03 & ~0.5\\
141937 & 77740 & 5855 & 4.37 & ~~1.2 & ~0.119 & ~~~2.21 & $<$1.9 & 1.05 & 4.40 & ~1.8\\
 & & ~~~20 & 0.03 & 0.06 & ~0.016 & ~~~0.05 & ~~0.5 & 0.03 & 0.03 & ~1.7\\
149661 & 81300 & 5184 & 4.41 & ~~0.8 & ~0.051 & ~~-0.20 & $<$2.2 & 0.86 & 4.53 & ~4.5\\
 & & ~~~39 & 0.06 & 0.08 & ~0.024 & ~~~0.05 & ~~0.3 & 0.02 & 0.02 & ~4.1\\
\hline
\end{tabular}
\end{minipage}
\end{table*}

\begin{table*}
\centering
\contcaption{}
\begin{tabular}{llrcccccccc}
\hline
Star & & T$_{\rm eff}$ & log g & $\zeta_{\rm t}$ (km~s$^{\rm -1}$) & [Fe/H] & log $\epsilon$(Li) & vsini (km~s$^{\rm -1}$) & mass (M$_{\odot}$) & log g & age (Gyr)\\
HD & HIP & $\pm$ & $\pm$ & $\pm$ & $\pm$ & $\pm$ & $\pm$ & $\pm$ & $\pm$ & $\pm$\\
\hline
152391 & 82588 & 5450 & 4.51 & ~~1.2 & ~0.003 & ~~~1.11 & ~~3.8 & 0.90 & 4.50 & ~3.9\\
 & & ~~~31 & 0.07 & 0.07 & ~0.022 & ~~~0.06 & ~~0.5 & 0.03 & 0.02 & ~3.5\\
165401 & 88622 & 5790 & 4.62 & ~~1.5 & -0.416 & ~~~0.75 & ~~4.5 & 0.86 & 4.42 & ~8.6\\
 & & ~~~31 & 0.05 & 0.12 & ~0.023 & ~~~0.07 & ~~0.5 & 0.02 & 0.03 & ~3.0\\
168443 & 89844 & 5524 & 3.99 & ~~1.0 & ~0.083 & ~~~0.43 & ~~2.5 & 1.01 & 4.03 & 10.1\\
 & & ~~~24 & 0.05 & 0.05 & ~0.018 & ~~~0.05 & ~~0.4 & 0.02 & 0.03 & ~0.6\\
168746 & 90004 & 5549 & 4.35 & ~~1.1 & -0.093 & ~~~0.72 & $<$0.5 & 0.91 & 4.36 & 10.1\\
 & & ~~~26 & 0.03 & 0.06 & ~0.022 & ~~~0.05 & ~~0.5 & 0.02 & 0.03 & ~1.9\\
170469 & 90593 & 5766 & 4.25 & ~~1.2 & ~0.306 & ~~~1.31 & $<$1.7 & 1.07 & 4.23 & 5.8\\
 & & ~~~31 & 0.03 & 0.04 & ~0.023 & ~~~0.12 & ~~0.8 & 0.03 & 0.06 & 1.9\\
190228 & 98714 & 5235 & 3.60 & ~~1.0 & -0.267 & ~~~1.19 & ~~2.0 & 1.17 & 3.68 & ~5.3\\
 & & ~~~22 & 0.05 & 0.09 & ~0.017 & ~~~0.05 & ~~0.6 & 0.04 & 0.04 & ~0.7\\
195019 & 100970 & 5780 & 4.15 & ~~1.3 & ~0.061 & ~~~1.27 & $<$2.0 & 1.05 & 4.10 & ~7.6\\
 & & ~~~19 & 0.02 & 0.03 & ~0.015 & ~~~0.05 & ~~0.6 & 0.02 & 0.04 & ~0.9\\
202206 & 104903 & 5709 & 4.39 & ~~1.1 & ~0.322 & ~~~1.40 & $<$1.0 & 1.02 & 4.42 & ~2.1\\
 & & ~~~26 & 0.03 & 0.05 & ~0.019 & ~~~0.05 & ~~0.5 & 0.03 & 0.03 & ~1.9\\
\hline
\end{tabular}
\end{table*}

\begin{table*}
\centering
\begin{minipage}{160mm}
\caption{Parameters determined for the comparison stars. Columns are the same as in Table 1.}
\label{xmm}
\begin{tabular}{llccccccccc}
\hline
Star & & T$_{\rm eff}$ & log g & $\zeta_{\rm t}$ (km~s$^{\rm -1}$) & [Fe/H] & log $\epsilon$(Li) & vsini (km~s$^{\rm -1}$) & mass (M$_{\odot}$) & log g & age (Gyr)\\
HD & HIP & $\pm$ & $\pm$ & $\pm$ & $\pm$ & $\pm$ & $\pm$ & $\pm$ & $\pm$ & $\pm$\\
\hline
S$^{4}$N & & & & & & & & & &\\
4614 & 3821 & 5904 & 4.32 & ~~1.2 & -0.246 & ~~~2.03 & $<$2.5 & 0.94 & 4.34 & 8.2\\
 & & ~~~32 & 0.03 & 0.11 & ~0.024 & ~~~0.06 & ~~0.5 & 0.03 & 0.03 & 2.4\\
9826 & 7513 & 6239 & 4.19 & ~~1.7 & ~0.145 & ~~~2.20 & ~~9.5 & 1.29 & 4.14 & 2.4\\
 & & ~~~37 & 0.03 & 0.06 & ~0.026 & ~~~0.06 & ~~0.4 & 0.02 & 0.03 & 0.3\\
10307 & 7918 & 5847 & 4.15 & ~~1.1 & ~0.041 & ~~~1.71 & $<$3.5 & 1.03 & 4.31 & 5.6\\
 & & ~~~25 & 0.04 & 0.07 & ~0.019 & ~~~0.06 & ~~0.7 & 0.03 & 0.04 & 2.4\\
13974 & 10644 & 5799 & 4.29 & ~~0.8 & -0.439 & ~~~2.21 & $<$1.5 & 0.87 & 4.35 & 10.\\
 & & ~~~70 & 0.09 & 0.50 & ~0.050 & ~~~0.08 & ~~0.1 & 0.01 & 0.02 & 1.1\\
16895 & 12777 & 6329 & 4.32 & ~~1.6 & ~0.057 & ~~~2.80 & ~~9.0 & 1.22 & 4.30 & 1.0\\
 & & ~~~37 & 0.02 & 0.08 & ~0.026 & ~~~0.06 & ~~0.5 & 0.02 & 0.02 & 0.8\\
17206 & 12843 & 6504 & 4.44 & ~~1.6 & ~0.279 & ~~~1.94 & ~~27. & 1.25 & 4.31 & 0.5\\
 & & ~~211 & 0.16 & 0.30 & ~0.150 & ~~~0.18 & ~~1.0 & 0.03 & 0.02 & 0.4\\
17925 & 13402 & 5060 & 4.25 & ~~1.1 & ~0.120 & ~~~2.27 & ~~6.0 & 0.85 & 4.53 & 4.8\\
 & & ~~~79 & 0.16 & 0.30 & ~0.040 & ~~~0.14 & ~~0.5 & 0.02 & 0.02 & 4.1\\
19373 & 14632 & 6008 & 4.33 & ~~1.3 & ~0.151 & ~~~2.27 & ~~4.5 & 1.16 & 4.21 & 3.8\\
 & & ~~~33 & 0.03 & 0.07 & ~0.023 & ~~~0.06 & ~~0.5 & 0.02 & 0.03 & 0.9\\
20010 & 14879 & 6170 & 3.93 & ~~1.7 & -0.206 & ~~~1.86 & ~~5.0 & 1.25 & 3.94 & 3.8\\
 & & ~~~35 & 0.02 & 0.09 & ~0.025 & ~~~0.06 & ~~0.6 & 0.04 & 0.03 & 0.7\\
20630 & 15457 & 5749 & 4.51 & ~~1.1 & ~0.078 & ~~~1.75 & ~~5.5 & 1.00 & 4.46 & 1.7\\
 & & ~~~26 & 0.05 & 0.05 & ~0.018 & ~~~0.06 & ~~0.5 & 0.03 & 0.02 & 1.7\\
22049 & 16537 & 5032 & 4.30 & ~~0.6 & -0.012 & ~~~0.05 & ~~2.5 & 0.82 & 4.57 & 3.2\\
 & & ~~~48 & 0.08 & 0.12 & ~0.021 & ~~~0.09 & ~~0.5 & 0.02 & 0.02 & 3.6\\
22484 & 16852 & 5996 & 3.97 & ~~1.4 & -0.045 & ~~~2.10 & ~~4.5 & 1.15 & 4.06 & 5.3\\
 & & ~~~26 & 0.01 & 0.06 & ~0.019 & ~~~0.05 & ~~0.8 & 0.04 & 0.03 & 1.0\\
30495 & 22263 & 5823 & 4.42 & ~~1.1 & ~0.042 & ~~~2.29 & ~~4.0 & 1.01 & 4.44 & 2.0\\
 & & ~~~26 & 0.05 & 0.07 & ~0.019 & ~~~0.05 & ~~0.5 & 0.03 & 0.02 & 1.8\\
34411 & 24813 & 5858 & 4.14 & ~~1.1 & ~0.103 & ~~~1.95 & $<$2.0 & 1.06 & 4.23 & 6.0\\
 & & ~~~19 & 0.03 & 0.09 & ~0.016 & ~~~0.05 & ~~0.5 & 0.03 & 0.03 & 1.3\\
39587 & 27913 & 5974 & 4.43 & ~~1.3 & ~0.018 & ~~~2.74 & ~~9.0 & 1.05 & 4.43 & 1.2\\
 & & ~~~33 & 0.05 & 0.06 & ~0.025 & ~~~0.06 & ~~0.5 & 0.03 & 0.02 & 1.1\\
61421 & 37279 & 6593 & 3.90 & ~~1.8 & ~0.022 & ~$<$0.81 & ~~5.0 & 1.47 & 3.99 & 1.8\\
 & & ~~~50 & 0.01 & 0.12 & ~0.032 & ~~~0.14 & ~~0.6 & 0.02 & 0.02 & 0.1\\
69830 & 40693 & 5382 & 4.34 & ~~0.8 & -0.018 & ~~~0.75 & ~~3.5 & 0.89 & 4.49 & 5.0\\
 & & ~~~25 & 0.09 & 0.12 & ~0.017 & ~~~0.06 & ~~0.5 & 0.03 & 0.03 & 4.2\\
72905 & 42438 & 5873 & 4.44 & ~~1.5 & -0.043 & ~~~2.65 & ~10.0 & 1.00 & 4.44 & 2.1\\
 & & ~~~34 & 0.05 & 0.07 & ~0.026 & ~~~0.06 & ~~0.6 & 0.03 & 0.02 & 1.9\\
82328 & 46853 & 6336 & 3.72 & ~~1.7 & -0.140 & ~~~3.09 & ~~8.5 & 1.44 & 3.84 & 2.4\\
 & & ~~~62 & 0.05 & 0.13 & ~0.042 & ~~~0.07 & ~~0.5 & 0.05 & 0.03 & 0.3\\
90839 & 51459 & 6146 & 4.37 & ~~1.5 & -0.061 & ~~~2.65 & $<$1.5 & 1.10 & 4.36 & 1.8\\
 & & ~~~33 & 0.02 & 0.15 & ~0.024 & ~~~0.06 & ~~0.8 & 0.03 & 0.03 & 1.5\\
95128 & 53721 & 5842 & 4.13 & ~~1.1 & ~0.025 & ~~~1.61 & $<$2.0 & 1.02 & 4.24 & 7.0\\
 & & ~~~31 & 0.05 & 0.09 & ~0.024 & ~~~0.06 & ~~0.5 & 0.03 & 0.04 & 1.6\\
102870 & 55757 & 6111 & 4.00 & ~~1.4 & ~0.160 & ~~~1.86 & ~~4.0 & 1.30 & 4.08 & 3.0\\
 & & ~~~28 & 0.05 & 0.06 & ~0.021 & ~~~0.05 & ~~0.5 & 0.03 & 0.03 & 0.4\\
109358 & 61317 & 5806 & 4.16 & ~~1.1 & -0.220 & ~~~1.50 & $<$2.5 & 0.92 & 4.32 & 10.\\
 & & ~~~24 & 0.07 & 0.08 & ~0.019 & ~~~0.05 & ~~0.5 & 0.02 & 0.03 & 1.7\\
114710 & 64394 & 6021 & 4.37 & ~~1.2 & ~0.083 & ~~~2.50 & ~~4.3 & 1.10 & 4.38 & 1.4\\
 & & ~~~21 & 0.05 & 0.04 & ~0.015 & ~~~0.05 & ~~0.7 & 0.03 & 0.02 & 1.2\\
115617 & 64924 & 5533 & 4.29 & ~~0.9 & ~0.010 & ~$<$0.10 & ~~2.2 & 0.93 & 4.42 & 6.8\\
 & & ~~~19 & 0.03 & 0.06 & ~0.013 & ~~~0.06 & ~~0.5 & 0.03 & 0.03 & 3.9\\
121370 & 67927 & 6085 & 3.70 & ~~1.7 & ~0.283 & ~~~1.57 & ~13.0 & 1.60 & 3.77 & 2.0\\
 & & ~~~54 & 0.10 & 0.07 & ~0.041 & ~~~0.07 & ~~0.5 & 0.03 & 0.20 & 0.1\\
131156 & 72659 & 5595 & 4.70 & ~~1.1 & -0.075 & ~~~2.24 & ~~5.0 & 0.90 & 4.50 & 3.3\\
 & & ~~~24 & 0.06 & 0.06 & ~0.015 & ~~~0.06 & ~~0.6 & 0.03 & 0.02 & 2.8\\
133640 & 73695 & 5848 & 4.32 & ~~0.9 & -0.186 & ~~~1.95 & $<$3.5 & 0.95 & 4.24 & 9.9\\
 & & ~~~33 & 0.03 & 0.11 & ~0.025 & ~~~0.06 & ~~0.5 & 0.02 & 0.03 & 1.2\\
 141004 & 77257 & 5869 & 4.00 & ~~1.1 & ~0.010 & ~~~1.73 & ~~3.2 & 1.05 & 4.15 & 7.1\\
 & & ~~~32 & 0.09 & 0.08 & ~0.025 & ~~~0.06 & ~~0.5 & 0.02 & 0.03 & 1.1\\
\hline
\end{tabular}
\end{minipage}
\end{table*}

\begin{table*}
\centering
\contcaption{}
\label{xmm}
\begin{tabular}{llccccccccc}
\hline
Star & & T$_{\rm eff}$ & log g & $\zeta_{\rm t}$ (km~s$^{\rm -1}$) & [Fe/H] & log $\epsilon$(Li) & vsini (km~s$^{\rm -1}$) & mass (M$_{\odot}$) & log g & age (Gyr)\\
HD & HIP & $\pm$ & $\pm$ & $\pm$ & $\pm$ & $\pm$ & $\pm$ & $\pm$ & $\pm$ & $\pm$\\
\hline
142860 & 78072 & 6209 & 3.91 & ~~1.4 & -0.160 & ~~~2.03 & ~11.0 & 1.15 & 4.12 & 4.9\\
 & & ~~~55 & 0.08 & 0.07 & ~0.037 & ~~~0.07 & ~~0.5 & 0.04 & 0.03 & 0.9\\
146233 & 79672 & 5753 & 4.22 & ~~1.0 & ~0.051 & ~~~1.43 & ~~2.7 & 1.00 & 4.39 & 3.9\\
 & & ~~~37 & 0.09 & 0.07 & ~0.030 & ~~~0.07 & ~~0.6 & 0.04 & 0.04 & 3.1\\
150680 & 81693 & 5827 & 3.79 & ~~1.4 & ~0.098 & ~$<$0.64 & ~~4.5 & 1.41 & 3.74 & 2.9\\
 & & ~~~18 & 0.03 & 0.05 & ~0.015 & ~~~0.07 & ~~0.5 & 0.03 & 0.03 & 0.2\\
157214 & 84862 & 5736 & 4.39 & ~~0.9 & -0.315 & ~$<$0.07 & ~~1.9 & 0.89 & 4.31 & 11.\\
 & & ~~~37 & 0.05 & 0.18 & ~0.027 & ~~~0.05 & ~~0.6 & 0.01 & 0.01 & 0.6\\
160269 & 86036 & 6026 & 4.58 & ~~1.2 & ~0.078 & ~~~2.50 & ~~5.0 & 1.10 & 4.39 & 1.3\\
 & & ~~~41 & 0.04 & 0.10 & ~0.030 & ~~~0.06 & ~~0.5 & 0.03 & 0.02 & 1.1\\
165896 & 88937 & 6206 & 4.38 & ~~1.8 & -0.525 & ~~~2.26 & ~~9.0 & 0.94 & 4.20 & 8.7\\
 & & ~~~85 & 0.05 & 0.60 & ~0.055 & ~~~0.08 & ~~0.5 & 0.03 & 0.03 & 1.3\\
222368 & 116771 & 6104 & 3.87 & ~~1.4 & -0.120 & ~~~2.09 & ~~7.5 & 1.17 & 4.05 & 4.9\\
 & & ~~~61 & 0.07 & 0.15 & ~0.050 & ~~~0.07 & ~~0.5 & 0.04 & 0.03 & 0.9\\
APR02 & & & & & & & & & &\\
69897 & 40843 & 6232 & 4.12 & ~~1.7 & -0.255 & ~~~2.57 & ~~5.7 & 1.08 & 4.18 & 5.5\\
 & & ~~~62 & 0.03 & 0.18 & ~0.041 & ~~~0.05 & ~~0.7 & 0.03 & 0.03 & 0.8\\
84737 & 48113 & 5934 & 4.16 & ~~1.3 & ~0.163 & ~~~2.28 & ~~3.0 & 1.17 & 4.13 & 4.5\\
 & & ~~~33 & 0.03 & 0.07 & ~0.026 & ~~~0.06 & ~~0.6 & 0.04 & 0.03 & 0.8\\
86728 & 49081 & 5753 & 4.30 & ~~1.1 & ~0.260 & ~~~1.27 & ~~3.3 & 1.05 & 4.32 & 4.7\\
 & & ~~~41 & 0.04 & 0.10 & ~0.030 & ~~~0.06 & ~~0.5 & 0.03 & 0.02 & 1.1\\
99491 & 55846 & 5415 & 4.37 & ~~1.1 & ~0.303 & ~~~0.83 & $<$1.4 & 0.94 & 4.44 & 4.8\\
 & & ~~~39 & 0.05 & 0.08 & ~0.027 & ~~~0.07 & ~~0.7 & 0.03 & 0.03 & 4.0\\
101501 & 56997 & 5530 & 4.46 & ~~1.0 & -0.002 & ~~~0.86 & $<$2.2 & 0.93 & 4.41 & 4.8\\
 & & ~~~24 & 0.03 & 0.08 & ~0.017 & ~~~0.05 & ~~0.4 & 0.06 & 0.15 & 4.2\\
102634 & 57629 & 6342 & 4.24 & ~~1.8 & ~0.219 & ~~~2.40 & ~~6.6 & 1.31 & 4.19 & 1.6\\
 & & ~~~59 & 0.03 & 0.11 & ~0.042 & ~~~0.05 & ~~0.4 & 0.02 & 0.03 & 0.5\\
102870 & 57757 & 6180 & 4.15 & ~~1.4 & ~0.213 & ~~~1.94 & ~~4.0 & 1.30 & 4.11 & 2.5\\
 & & ~~~36 & 0.05 & 0.07 & ~0.026 & ~~~0.06 & ~~0.6 & 0.01 & 0.02 & 0.3\\
109358 & 61317 & 5887 & 4.34 & ~~1.5 & -0.198 & ~~~1.63 & $<$2.5 & 0.95 & 4.33 & 8.1\\
 & & ~~~26 & 0.03 & 0.11 & ~0.021 & ~~~0.05 & ~~0.6 & 0.03 & 0.04 & 2.7\\
114710 & 64394 & 6020 & 4.35 & ~~1.3 & ~0.083 & ~~~2.40 & ~~4.4 & 1.10 & 4.37 & 1.5\\
 & & ~~~28 & 0.02 & 0.06 & ~0.020 & ~~~0.05 & ~~0.6 & 0.03 & 0.03 & 1.3\\
115383 & 64792 & 6100 & 4.28 & ~~1.5 & ~0.192 & ~~~2.69 & ~~7.1 & 1.19 & 4.27 & 2.0\\
 & & ~~~35 & 0.06 & 0.05 & ~0.026 & ~~~0.06 & ~~0.3 & 0.02 & 0.03 & 1.1\\
120066 & 67246 & 5862 & 4.12 & ~~1.4 & ~0.081 & ~~~2.67 & ~~3.5 & 1.11 & 4.11 & 6.0\\
 & & ~~~25 & 0.03 & 0.06 & ~0.020 & ~~~0.05 & ~~0.5 & 0.03 & 0.04 & 1.0\\
122652 & 68593 & 6199 & 4.46 & ~~1.4 & ~0.063 & ~~~2.69 & ~~4.2 & 1.15 & 4.36 & 0.9\\
 & & ~~~34 & 0.04 & 0.08 & ~0.025 & ~~~0.05 & ~~0.5 & 0.02 & 0.02 & 0.7\\
134044 & 73941 & 6290 & 4.48 & ~~1.6 & ~0.131 & ~~~2.70 & ~~4.4 & 1.21 & 4.32 & 0.7\\
 & & ~~~38 & 0.05 & 0.08 & ~0.026 & ~~~0.05 & ~~0.6 & 0.02 & 0.02 & 0.6\\
135101A & 74432 & 5665 & 4.14 & ~~1.3 & ~0.054 & ~~~0.95 & $<$1.8 & 0.90 & 4.27 & 9.2\\
 & & ~~~24 & 0.04 & 0.04 & ~0.019 & ~~~0.11 & ~~0.7 & 0.02 & 0.04 & 1.9\\
145825 & 79578 & 5795 & 4.43 & ~~1.2 & ~0.054 & ~~~1.85 & $<$1.4 & 1.01 & 4.44 & 1.9\\
 & & ~~~25 & 0.03 & 0.06 & ~0.019 & ~~~0.05 & ~~1.0 & 0.03 & 0.03 & 1.8\\
147044 & 79862 & 5852 & 4.42 & ~~1.2 & -0.016 & ~~~2.06 & $<$2.0 & 1.01 & 4.37 & 3.0\\
 & & ~~~41 & 0.08 & 0.10 & ~0.030 & ~~~0.06 & ~~0.8 & 0.04 & 0.04 & 3.0\\
150433 & 81681 & 5678 & 4.45 & ~~1.2 & -0.326 & ~~~0.91 & $<$1.2 & 0.87 & 4.39 & 10.\\
 & & ~~~32 & 0.07 & 0.11 & ~0.025 & ~~~0.10 & ~~0.7 & 0.02 & 0.02 & 1.4\\
153458 & 83181 & 5841 & 4.51 & ~~1.2 & ~0.133 & ~~~2.14 & $<$0.5 & 1.05 & 4.40 & 1.7\\
 & & ~~~21 & 0.03 & 0.07 & ~0.016 & ~~~0.06 & ~~0.8 & 0.03 & 0.03 & 1.7\\
157347 & 85042 & 5709 & 4.48 & ~~1.1 & ~0.071 & ~~~1.01 & $<$0.9 & 0.99 & 4.40 & 4.2\\
 & & ~~~26 & 0.05 & 0.07 & ~0.019 & ~~~0.05 & ~~0.9 & 0.04 & 0.04 & 3.3\\
159222 & 85810 & 5851 & 4.41 & ~~1.3 & ~0.155 & ~~~1.93 & ~~3.3 & 1.07 & 4.39 & 1.9\\
 & & ~~~20 & 0.02 & 0.06 & ~0.016 & ~~~0.06 & ~~0.8 & 0.03 & 0.03 & 1.7\\
161555 & 86985 & 5817 & 4.02 & ~~1.4 & ~0.116 & ~~~2.44 & $<$2.9 & 1.15 & 4.04 & 5.5\\
 & & ~~~38 & 0.05 & 0.06 & ~0.030 & ~~~0.05 & ~~0.4 & 0.04 & 0.05 & 1.0\\
162826 & 87382 & 6219 & 4.38 & ~~1.5 & ~0.075 & ~~~2.55 & ~~3.6 & 1.19 & 4.28 & 1.7\\
 & & ~~~37 & 0.03 & 0.07 & ~0.026 & ~~~0.05 & ~~0.5 & 0.03 & 0.03 & 1.1\\
\hline
\end{tabular}
\end{table*}

\begin{table*}
\centering
\contcaption{}
\label{xmm}
\begin{tabular}{llccccccccc}
\hline
Star & & T$_{\rm eff}$ & log g & $\zeta_{\rm t}$ (km~s$^{\rm -1}$) & [Fe/H] & log $\epsilon$(Li) & vsini (km~s$^{\rm -1}$) & mass (M$_{\odot}$) & log g & age (Gyr)\\
HD & HIP & $\pm$ & $\pm$ & $\pm$ & $\pm$ & $\pm$ & $\pm$ & $\pm$ & $\pm$ & $\pm$\\
\hline
164595 & 88194 & 5698 & 4.30 & ~~1.1 & -0.065 & ~~~0.96 & $<$0.5 & 0.94 & 4.38 & 6.8\\
 & & ~~~25 & 0.06 & 0.06 & ~0.019 & ~~~0.09 & ~~0.3 & 0.04 & 0.04 & 4.0\\
171067 & 90864 & 5644 & 4.53 & ~~1.1 & -0.015 & ~~~0.73 & $<$0.7 & 0.95 & 4.46 & 3.4\\
 & & ~~~21 & 0.05 & 0.10 & ~0.017 & ~~~0.11 & ~~0.7 & 0.03 & 0.03 & 3.2\\
\hline
\end{tabular}
\end{table*}

For the 31 SWPs in common between the present SWP sample and our previous studies, the mean difference in T$_{\rm eff}$ is only 15 K, while the mean $\sigma$ for T$_{\rm eff}$ is reduced from 43 K to 35 K. This demonstrates that our revised procedure for measuring EWs gives more precise T$_{\rm eff}$ values than the manual procedure we had employed in our previous studies, and it doesn't introduce significant systematic error.

Originally, the SWP sample in the present study consisted of 90 stars. However, for 5 of these stars (HD 33564, HD 114762, HD 136118, HD 141937 and HD 168443) the minimum masses of the plants are close to the brown dwarf limit (11 M$_{\rm J}$), and in some cases there is additional evidence that the true planet masses are beyond this limit. For these reasons, we transferred these stars to the comparison sample, leaving 85 stars in the SWPs sample.

We also determined age, mass and log g for each star from stellar isochrones. In particular, we employed our T$_{\rm eff}$ and [Fe/H] values with M$_{\rm v}$ calculated from the new reduction of the {\it Hipparcos} parallaxes \citep{van07} with a Bayesian parameter estimation method \citep{da06}.\footnote{We used Leo Girardi's web program PARAM to calculate these quantities. See: http://stev.oapd.inaf.it/cgi-bin/param.} The mean difference between our spectroscopic log g values and the parallax derived (photometric) values is -0.02 $\pm$ 0.12 dex. However, as can be seen in Figure 2 there is a significant trend present among the SWPs in the sense that the spectroscopic values are smaller than the isochrone ones; the trend is not significant if we exclude the cooler stars. If we restrict the comparison to T$_{\rm eff} > 5650$ K, then the difference is 0.00 $\pm$ 0.10 dex. This gives us confidence that the analysis results do not have systematic errors more than a few hundredths of a dex for this temperature range.

We show the Li abundances in Figure 3. The dividing/cutoff line between upper limits and detections of Li is about half a dex lower in the present sample compared to that in our previous study \citep{gg08}. We positioned the dividing line such that most stars with upper limits fall below it, and most stars with detections fall above it.

The number of comparison stars is very limited below about T$_{\rm eff} = 5650$ K, and the hottest SWP has a T$_{\rm eff}$ value just under 6350 K. Given this, we limit our comparison of Li abundances between SWPs and stars without planets to T$_{\rm eff} =$ 5650 to 6350 K. This compares to a T$_{\rm eff}$ range of 5550 to 6250 K in \citet{gg08}. In addition, for this choice of T$_{\rm eff}$ all the SWP and comparison stars fall above the Li dividing/cutoff line.

\begin{figure}
  \includegraphics[width=3.5in]{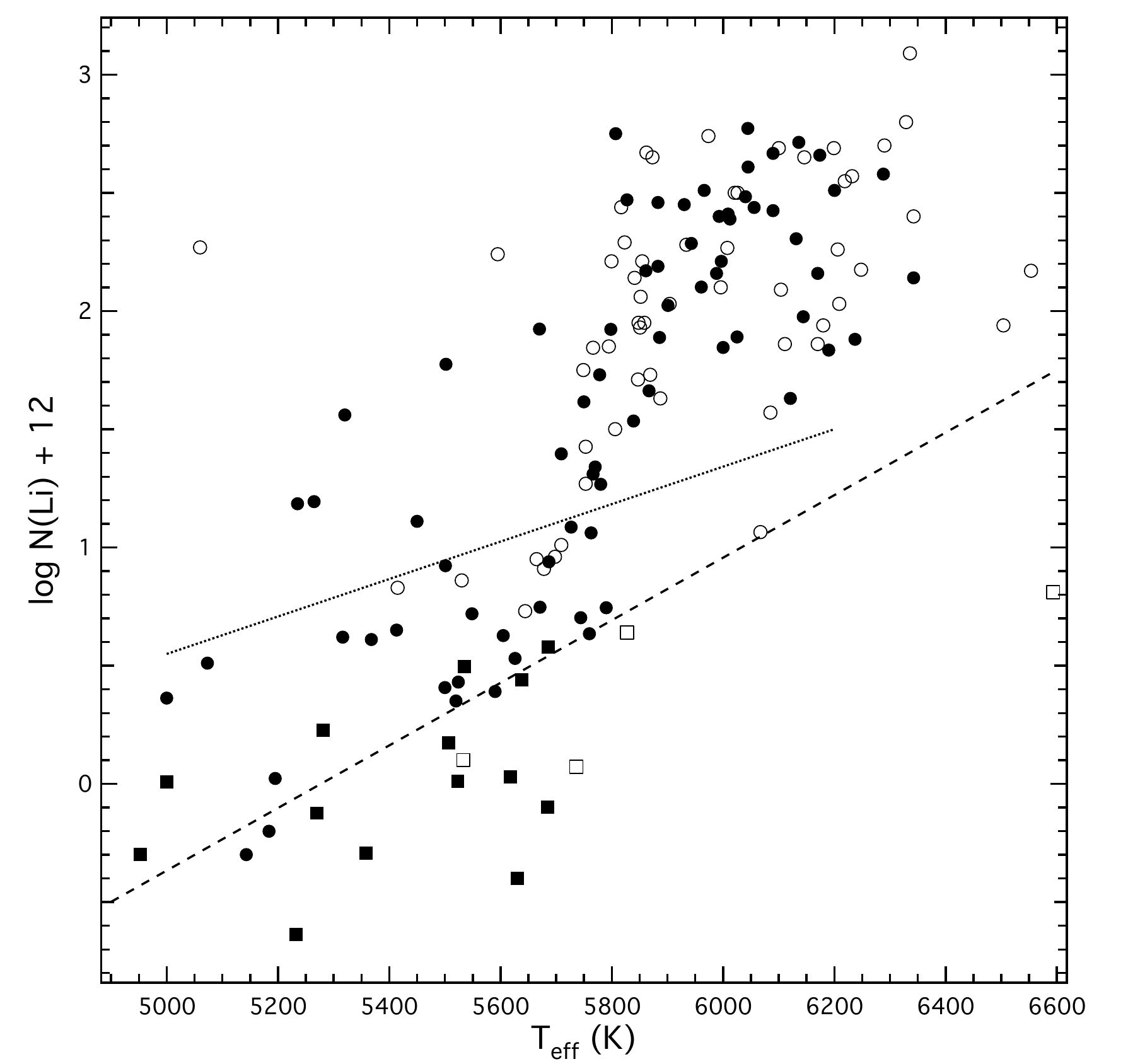}
 \caption{Li abundances versus T$_{\rm eff}$ for SWPs (dots) and stars without planets (open circles).  Upper limits on the Li abundance are shown for SWPs (filled squares) and stars without planets (open squares). The upper limit cutoff for the present data set is shown as a dashed line; the corresponding cutoff from \citet{gg08} is shown as a dotted line.}
\end{figure}

We plot our vsini estimates against T$_{\rm eff}$ in Figure 4. In this case, the stars with detected vsini are mixed over a large range in vsini with stars that have upper limits. For this reason, we set our vsini cutoff line rather high for the next step in the analysis. We also prepared a separate dataset with the vsini estimates of \citet{vf05} in place of our vsini estimates; we plot these data in Figure 5.\footnote{The lowest value of vsini quoted by \citet{vf05} is 0.0 km~s$^{\rm -1}$. In \citet{gg08} we reset stars from \citet{vf05} with a vsini value of 0.0 km~s$^{\rm -1}$ to 0.3 km~s$^{\rm -1}$. We do the same in the present work with their data. In addition, we set the minimum value of our vsini estimates to 0.5 km~s$^{\rm -1}$.}

For the 71 stars in the present study with vsini detections that overlap with the sample of \citet{vf05}, the mean difference in vsini is 0.1 $\pm$ 0.5 km~s$^{\rm -1}$.

\begin{figure}
  \includegraphics[width=3.5in]{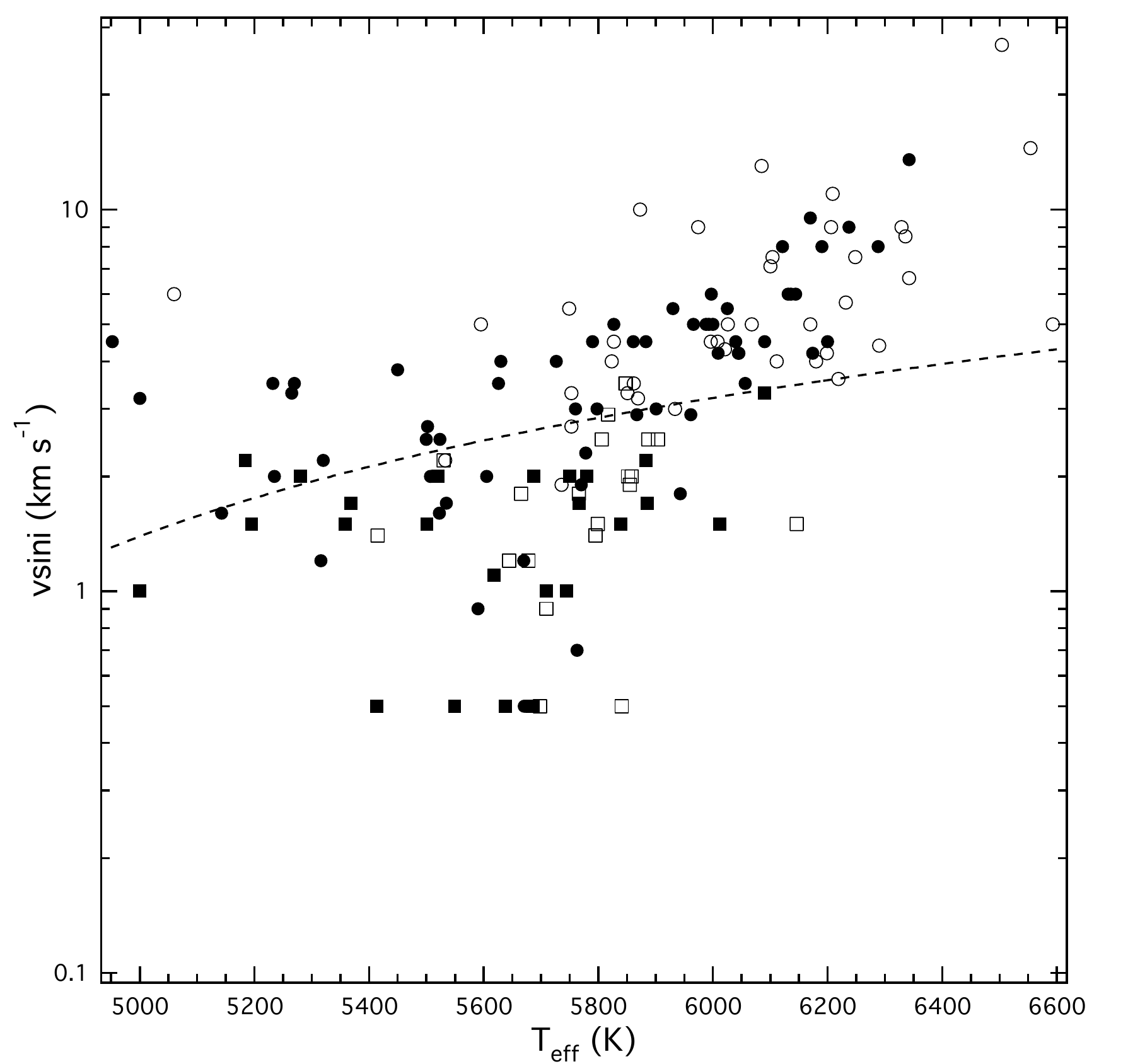}
 \caption{vsini versus T$_{\rm eff}$; symbols have the same meanings as in Figure 3. The upper limit cutoff is shown as a dashed line curve in this plot, but it is actually linear.}
\end{figure}

\begin{figure}
  \includegraphics[width=3.5in]{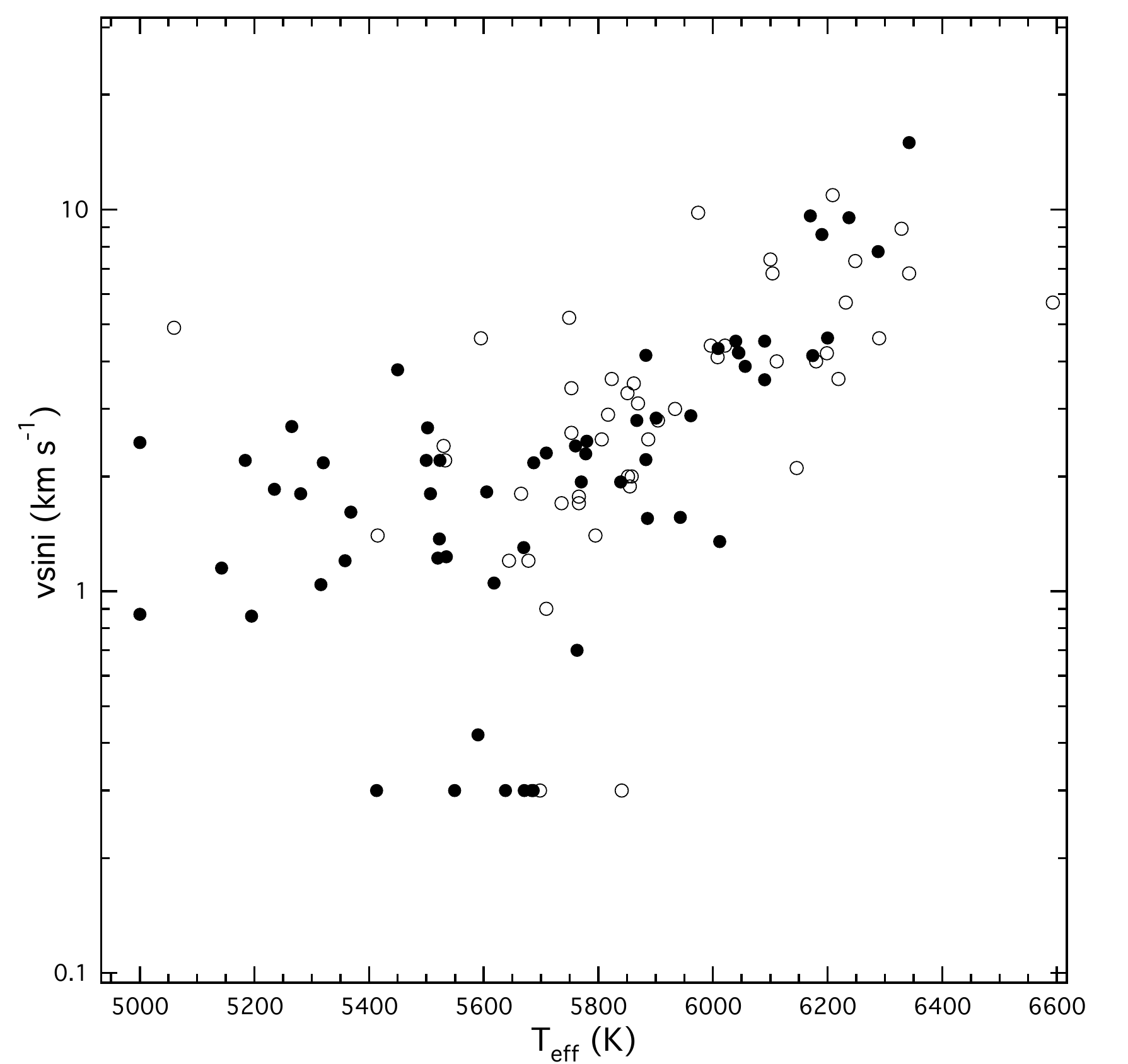}
 \caption{Same as Figure 4 but with vsini values from \citet{vf05}.}
\end{figure}

\section{Comparison of Samples}

In \citet{gg08} we introduced a new index, $\Delta_1$, which is a measure of the distance between two stars in T$_{\rm eff}$-[Fe/H]-log g-M$_{\rm v}$ space. We calculated a weighted average Li abundance difference between a given SWP and all the comparison stars with $(\Delta_1)^{-2}$ as the weight. We employ the same analysis here with our new samples.

One star in our SWP sample, HD 70573, lacks a {\it Hipparcos} parallax. Since the $\Delta_1$ index weighting scheme requires M$_{\rm v}$, we removed this star from our sample at this stage of our analysis. The number of retained SWPs in the range T$_{\rm eff} =$ 5650 to 6350 K is 50. In our comparison sample, 50 stars fall in this T$_{\rm eff}$ range, but one star has an upper limit on its Li abundance. We removed this star, leaving 49 stars in our comparison sample. This compares to 37 SWPs and 147 comparison stars over the range T$_{\rm eff} =$ 5550 to 6250 K in \citet{gg08}. We show the resulting weighted average Li abundance differences between the SWPs and comparison stars in Figure 6.

\begin{figure}
  \includegraphics[width=3.5in]{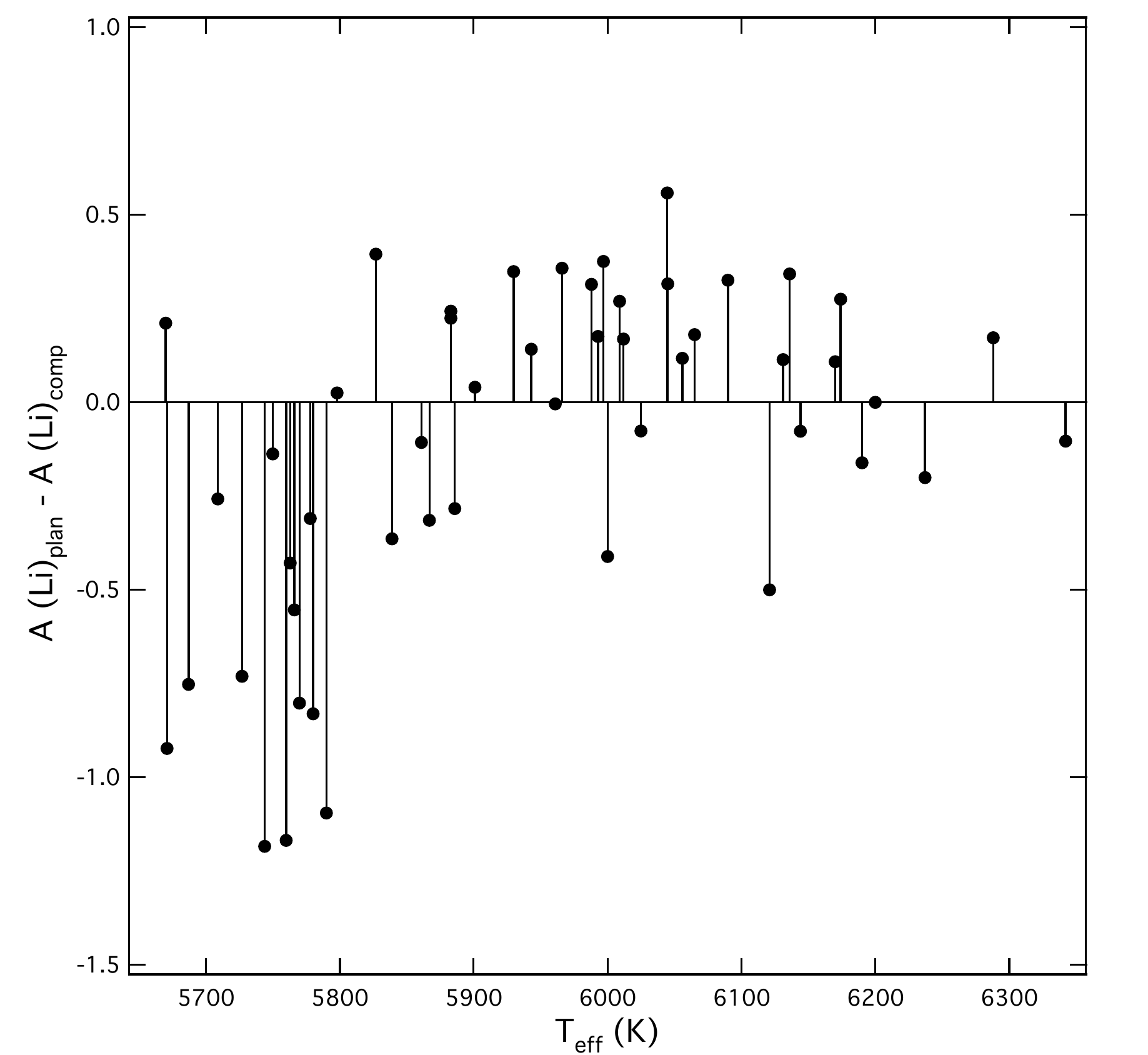}
 \caption{Weighted average Li abundance differences between SWPs and comparison stars analyzed in the present study.}
\end{figure}

\subsection{Correcting for bias in Li}

In order to test for possible bias in our method of sample comparison, we will revisit the Li abundance data from \citet{gg08}. If the Li abundance deficits we have detected among the SWPs near solar temperature are real, then there should be no trend with temperature when the comparison stars are analyzed in the same way. To realize this test, we selected every other star from the 147 comparison stars in \citet{gg08} and treated them as if they were SWPs (''fake SWPs''), and we treated the remaining stars as comparison stars. We then calculated the weighted Li abundance differences as before (Figure 7a). Next, we exchanged the roles of the stars and repeated the analysis (Figure 7b). There is a clear trend evident in these results; the slopes of least-squares fits to the two datasets in Figure 7 are 6.4$\times10^{\rm -4}$ and 8.9$\times10^{\rm -4}$ dex K$^{\rm -1}$, respectively. We adopt an average ''bias slope'' value of 7.7$\times10^{\rm -4}$ dex K$^{\rm -1}$.

We plot in Figure 8a the weighted Li abundance differences for the SWPs from \citet{gg08}, which differ from the data plotted in Figure 3b of that paper only in that the newly revised Hipparcos parallaxes have been used to calculate the $\Delta_1$ index instead of the outdated 1997 values. Next, we corrected these data for bias by subtracting the average bias slope value quoted above. The resulting data (Figure 8b) no longer display Li abundance excesses above $\sim6000$ K, but the Li abundance deficits are still present between about 5800 and 5900 K; there are not enough SWP data below 5800 K to reach a conclusion in this temperature range.

We repeated these tests with the datasets in the present study and show the results in Figures 9 and 10.
The slopes of the data in Figure 9 are 1.3$\times10^{\rm -3}$ and 4.4$\times10^{\rm -4}$ dex K$^{\rm -1}$. We adopt an average bias slope value of 8.7$\times10^{\rm -4}$ dex K$^{\rm -1}$. After applying this bias correction, we find that stars near the solar temperature still display the largest Li abundance deficits, but stars at intermediate temperatures also display modest Li abundance excesses.

There are 12 SWPs and 7 comparison stars between 5700 and 5800 K in the present study. If we apply a simple statistical t hypothesis test to the Li abundances from these samples (keeping in mind that we are not correcting for differences in the parameters here), we find that we can reject the hypothesis that these two samples are drawn from the same parent population with about 93\% confidence.

\begin{figure}
  \includegraphics[width=3.5in]{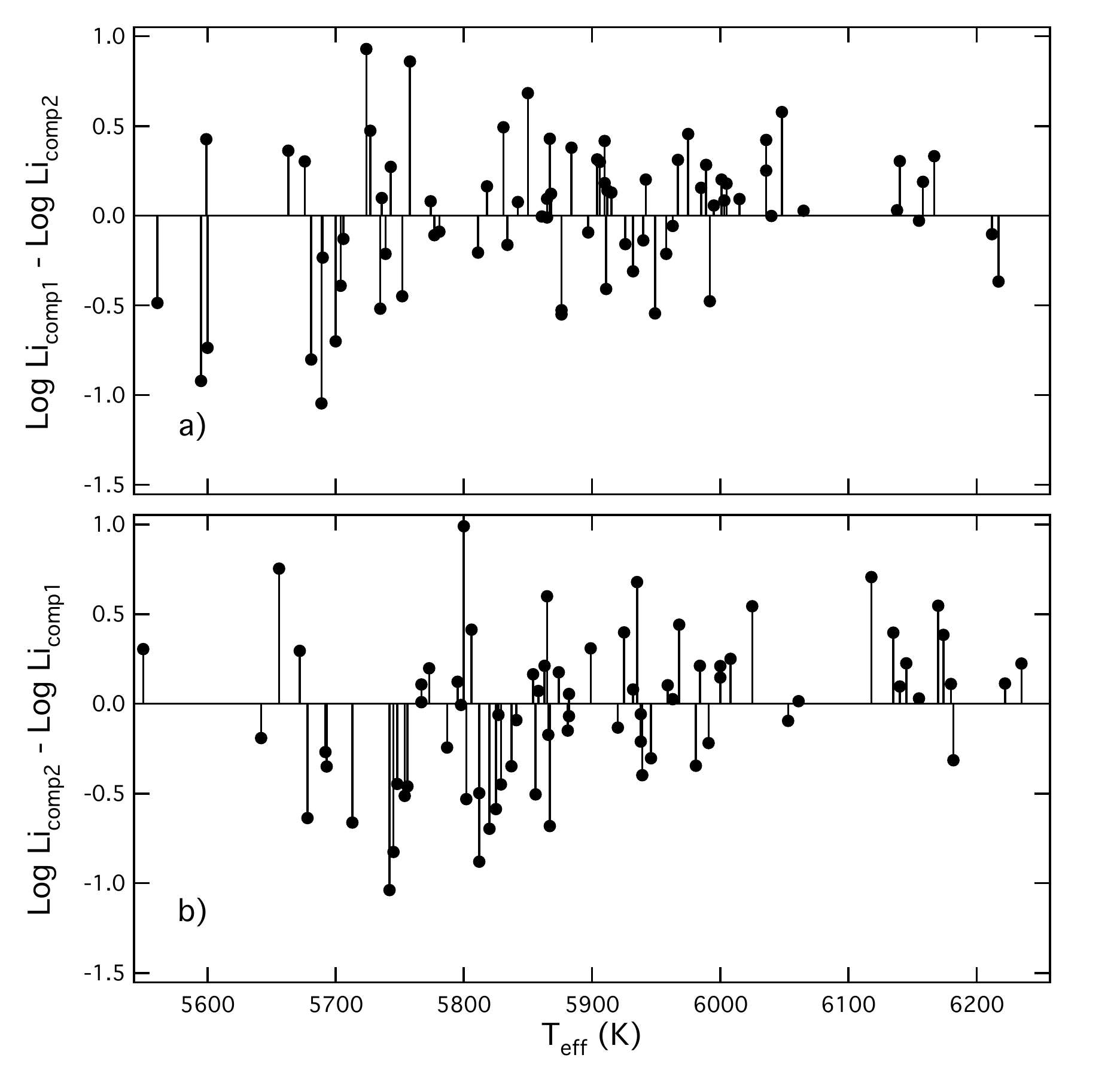}
 \caption{Weighted average Li abundance differences among the comparison stars from \citet{gg08}. Roles of the comparison stars have been exchanged (panel b). See text for details.}
\end{figure}

\begin{figure}
  \includegraphics[width=3.5in]{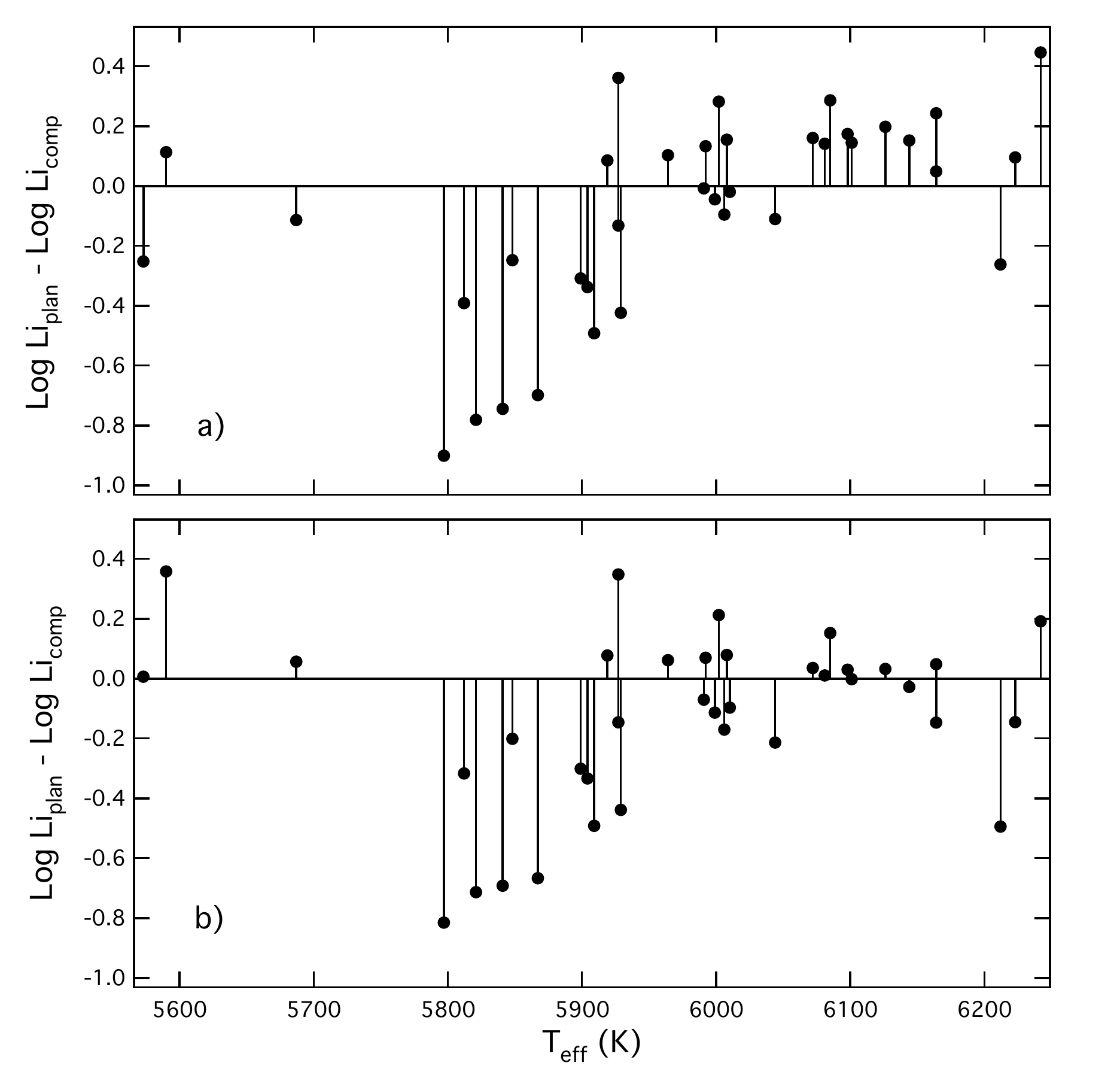}
 \caption{Weighted average Li abundance differences between SWPs and comparison stars for stars from \citet{gg08} (panel a). Li abundances corrected using the trends in Figure 7 (panel b).}
\end{figure}

\begin{figure}
  \includegraphics[width=3.5in]{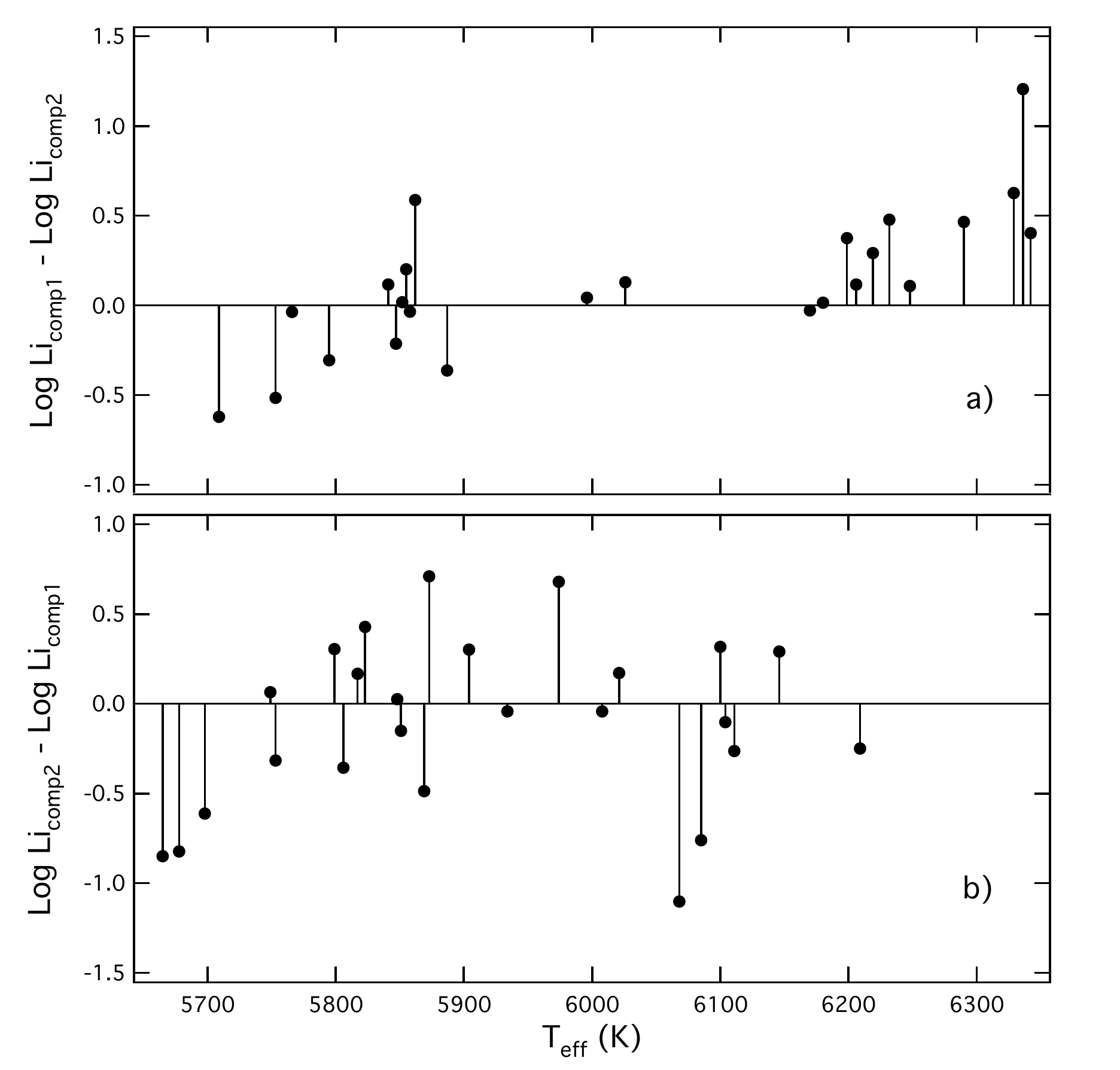}
 \caption{Same as Figure 7 but for comparison stars in the present study.}
\end{figure}

\begin{figure}
  \includegraphics[width=3.5in]{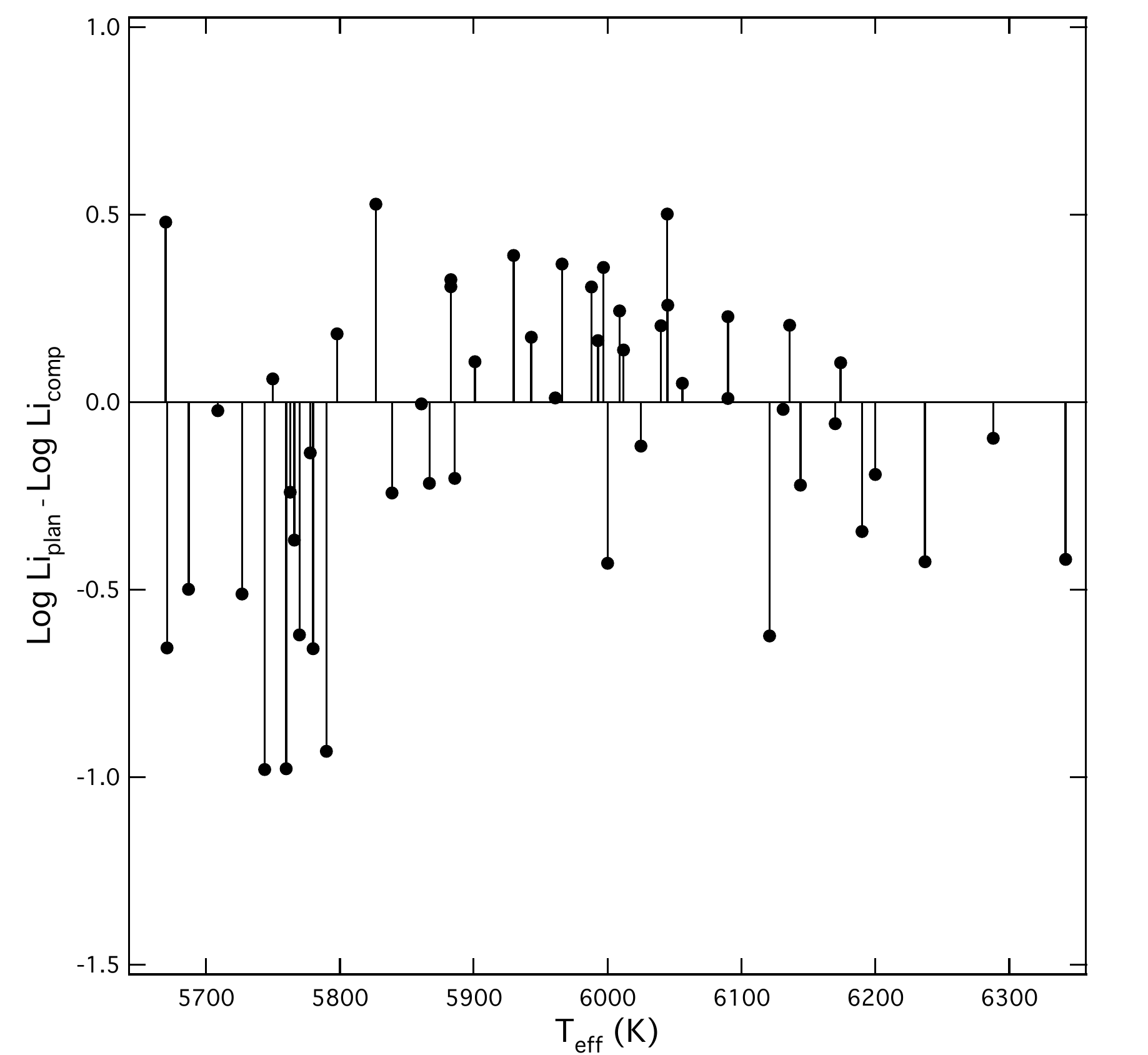}
 \caption{Same data as shown in Figure 6 but corrected for bias using average trend determined from data in Figure 9.}
\end{figure}

\begin{figure}
  \includegraphics[width=3.5in]{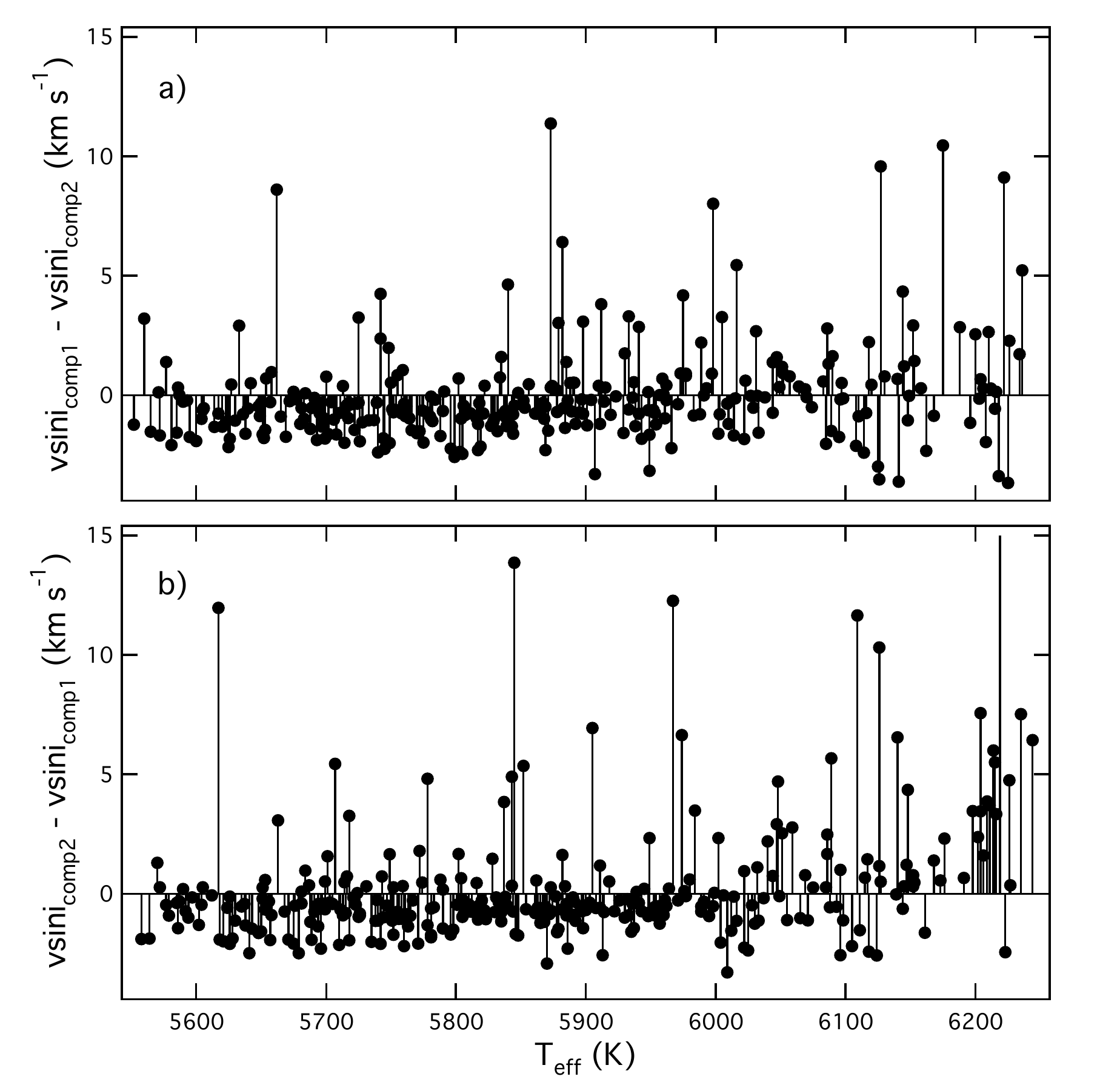}
 \caption{Weighted average vsini differences between two sets of comparison stars (like Figure 7). The vsini estimates are from \citet{vf05}.}
\end{figure}

\begin{figure}
  \includegraphics[width=3.5in]{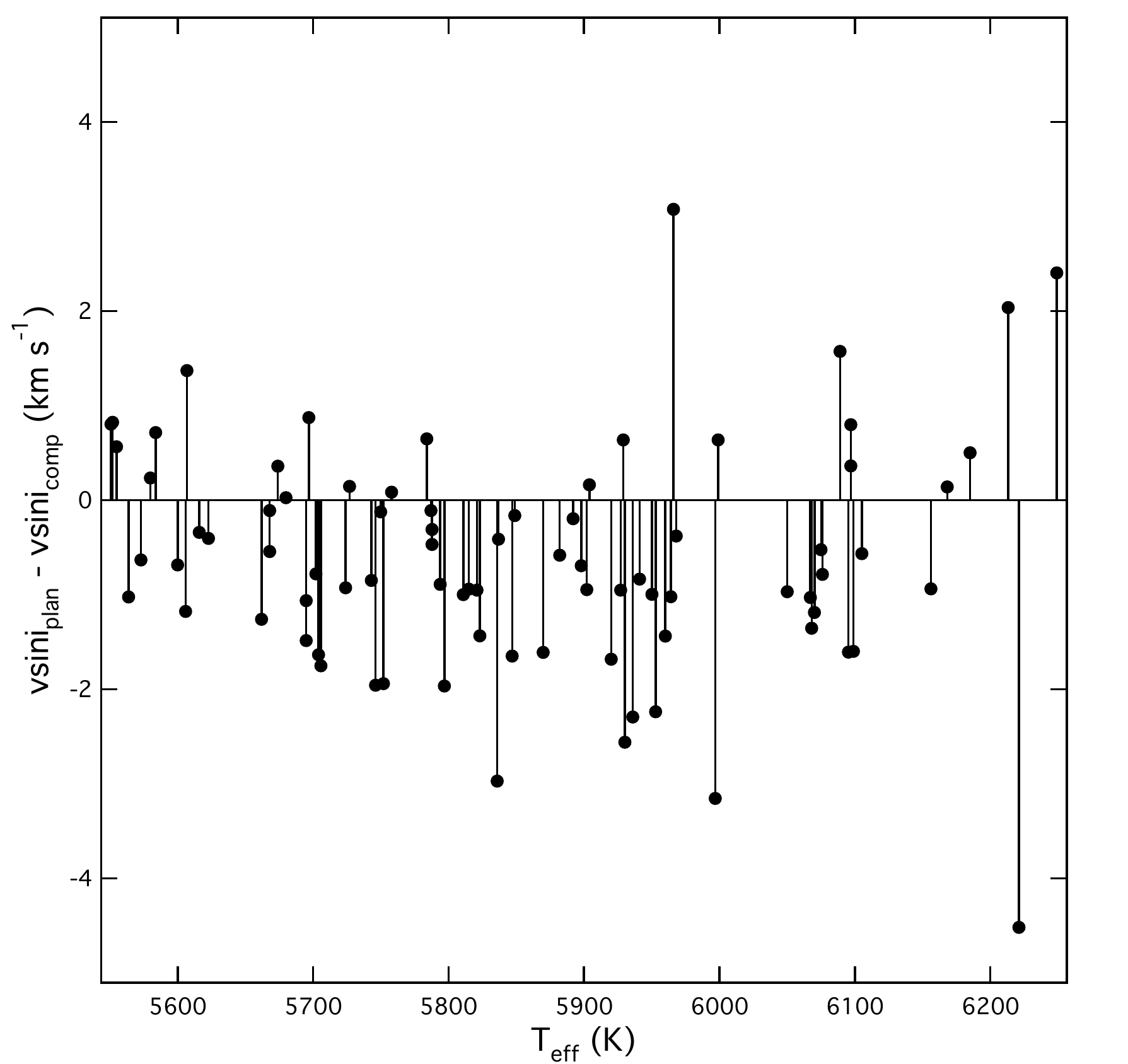}
 \caption{Corrected weighted average vsini differences between SWPs and comparison stars from \citet{vf05}. The corrections have been applied from the trends in Figure 11.}
\end{figure}

\subsection{Correcting for bias in vsini}

We applied an analysis to the vsini data very similar to our analysis of the Li abundances in the last section. First, we show in Figure 11 the weighted vsini differences for the 596 comparison stars with vsini measurements in \citet{vf05}. As for the Li abundance differences, there is a trend present in the vsini differences; the slopes are 0.0055 and 0.0026 km s$^{\rm -1}$ K$^{\rm -1}$, respectively. We adopt an average value of 0.0041 km s$^{\rm -1}$ K$^{\rm -1}$. The corrected SWP weighted vsini differences are shown in Figure 12. The vsini deficits among the SWPs are evident in the plot between about 5700 and 5900 K.

Due to the small number of comparison stars with vsini determinations in the present study (33), we are not confident that we can derive a reliable bias trend from them. For this reason, we have adopted the mean bias trend from Figure 11 and applied it to the 40 SWP vsini differences based on the new vsini determinations in the present study (Figure 13a). In applying the bias correction to the vsini differences, we shifted the temperature scale of the \citet{vf05} data to more closely match our temperature scale. The corresponding vsini differences for the 34 SWP \citet{vf05} vsini values of the stars included in the present study are shown in Figure 13b. vsini deficits are evident in both plots for temperatures between about 5700 and 5900 K.

\begin{figure}
  \includegraphics[width=3.5in]{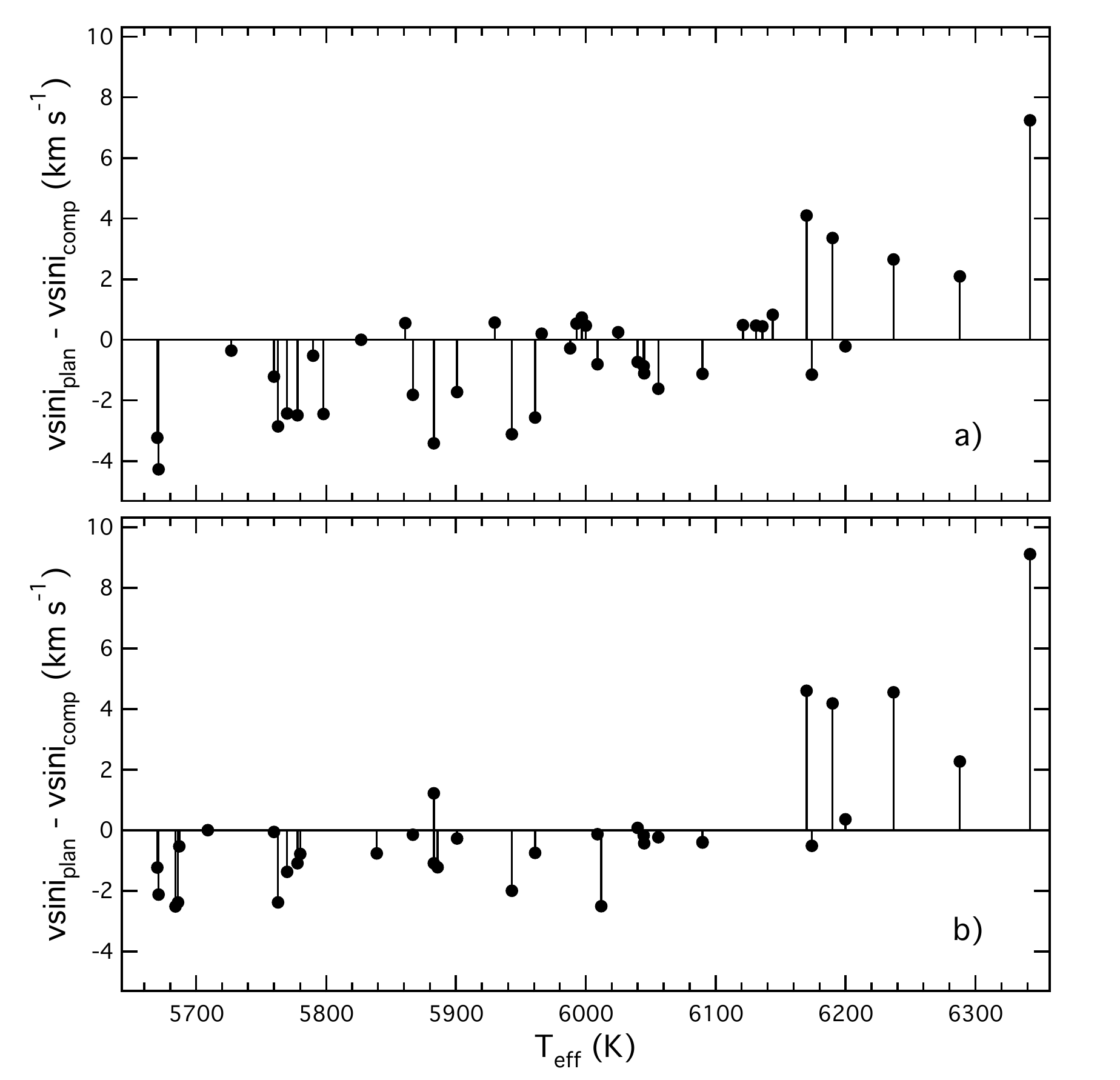}
 \caption{Weighted average vsini differences between SWPs and comparison stars versus T$_{\rm eff}$. Panel (a) is based on vsini estimates in the present study. Panel (b) is based on vsini estimates from \citet{vf05}.}
\end{figure}

\section{Discussion}

Despite the presence of an unrecognized bias in the results presented in \citet{gg08}, the general trend of weighted Li abundance differences with T$_{\rm eff}$ we determine is similar to the trend we presented in Figure 3b of that study. However, the data used in the present study show a rapid rise in Li abundance for SWPs with T$_{\rm eff} > 5850$, which is about 50 K cooler than the data in \citet{gg08}.  This is due to the fact that the T$_{\rm eff}$ scale adopted in \citet{gg08} is about 50 K higher than that adopted here. We consider the T$_{\rm eff}$ scale in the present study to be more accurate.

The trends of weighted vsini differences with T$_{\rm eff}$ are also similar to the trend we described in \citet{gg08}. In that study, we found that vsini transitions between deficits and excesses just below T$_{\rm eff} =$ 6100. In the present study we find that the transition occurs near 6000 K according to our vsini estimates and near 6100 K from the data of \citet{vf05}. This transition is not well constrained by the small number of SWPs with vsini estimates in the present study. Note, also, that the vsini values include a random component due to our ignorance of the inclination of a star's rotation axis. For this reason, we need a larger sample to study the trend with vsini than we employed for the Li abundance analysis.

One side benefit of the samples prepared for this study is that we can now fairly compare the Sun's Li abundance ($= 0.96$ dex) to other stars. The Li abundance cutoff employed in \citet{gg08} was such that the Sun would have been excluded from the SWP sample due to its low Li abundance. Now we are prepared to answer the question, ``Does the Sun's Li abundance better fit the classification of a SWP or of a star without a planet, according to the distribution in Figure 10?'' Applying the same analysis of the previous section to the Sun, we calculate a weighted solar Li abundance deficit of 0.70 dex relative to the 50 comparison stars. The average deficit for SWPs between 5700 and 5800 K This places the Sun squarely in the SWP category. Only 3 SWPs in Figure 10 have greater Li deficits than the Sun: HD 6434, HD 20782 and HD 165401.

The weighted solar vsini deficit is 1.25 km~s$^{\rm -1}$ relative to the larger \citet{vf05} vsini comparison star dataset.\footnote{We adopted a solar vsini value of 1.28 km~s$^{\rm -1}$, which is the average value a distant observer would observe, given a random distribution of possible inclinations of the rotation axis.} The average vsini deficit for the SWPs with T$_{\rm eff}$ between 5700 and 6000 K is 0.9 $\pm$ 1.1 km~s$^{\rm -1}$. This also places the Sun squarely in the SWP category.

\section{Conclusions}

We confirm our previous findings from \citet{gg08} that the Li abundances of SWPs with T$_{\rm eff} \leq 5800$ K are smaller than those of stars without detected planets; we also confirm that SWPs near $\sim 6000$ K have excess Li abundances. In addition, we confirm that SWPs have smaller vsini values than stars without detected planets for T$_{\rm eff}$ less than $\sim 6000$ K. These trends are robust, given that we employed different samples of SWPs and comparison stars in the two studies.

For the first time, we are able to compare the Sun's Li abundance in a fair way to a sample of comparison stars. We find that its Li abundance is low compared to our sample of comparison stars, and it is comparable to the Li abundances of SWPs with similar T$_{\rm eff}$ values.

To improve on our analysis, it is important to determine Li abundances and vsini values for more SWPs and comparison stars with T$_{\rm eff} > 5500$ K. To be most useful, spectra should have S/N ratios $> 300$ near 6700 \AA.

\section*{Acknowledgments}

We thank Peter Stetson for providing us with the most recent version of his DAOSPEC software; it saved us many hours of tedious measurements. We also thank Chris Laws for assistance with the April 2002 observations. The anonymous reviewer provided helpful suggestions that resulted in an improved paper. We acknowledge financial support from the Discovery Institute in Seattle, WA and Grove City College. This research has made use of the SIMBAD database, operated at CDS, Strasbourg, France.

\bsp

\label{lastpage}

\end{document}